\documentclass[12pt, a4paper]{article}

\usepackage{geometry}
\usepackage{verbatim}
\usepackage{graphicx}
\usepackage{subfigure}
\usepackage{overpic}
\usepackage{float}
\usepackage{amsmath}
\usepackage{bm}
\usepackage{txfonts}
\usepackage{array}
\usepackage{cite}

\newcommand{\mbB}{\bm{\mathrm{B}}}
\newcommand{\mbJ}{\bm{\mathrm{J}}}
\newcommand{\mbE}{\bm{\mathrm{E}}}
\newcommand{\mbV}{\bm{\mathrm{V}}}

\newcommand{\tg}{\text{g}}
\newcommand{\dif}{\mathrm{d}}
\newcommand{\mm}{\mathrm{m}}

\newcommand{\mT}{\mathrm{T}}
\newcommand{\mkeV}{\mathrm{keV}}

\newcommand{\efit}{\textsc{EFIT}}
\newcommand{\nimrod}{\textsc{NIMROD}}

\begin{document}
\begin{center}
  \Large\textbf{Transition from fishbone mode to
    $\beta$-induced Alfv\'en eigenmode on HL-2A tokamak}
\end{center}
Zhihui Zou$^{1}$, Ping Zhu$^{2, 3\ast}$, Charlson C. Kim$^{4}$, Xianqu Wang$^{5}$, Yawei Hou$^{1\ast\ast}$ 
\\
{\small
  $^{1}$ CAS Key Laboratory of Geospace Environment and Department of
  Plasma Physics and Fusion Engineering,
  University of Science and Technology of China, Hefei,
  Anhui 230026, China \\
  $^{2}$ International Joint Research Laboratory of Magnetic Confinement
  Fusion and Plasma Physics, State Key Laboratory of Advanced
  Electromagnetic Engineering and Technology, School of Electrical
  and Electronic Engineering, Huazhong University of Science and
  Technology, Wuhan, Hubei 430074, China\\
  $^{3}$ Department of Engineering Physics, University of
  Wisconsin-Madison, Madison, Wisconsin 53706,USA \\
  $^{4}$ SLS2 Consulting, San Diego, California 92107, USA \\
  $^{5}$ Institute of Fusion Science, School of Physical Science and
  Technology, Southwest Jiaotong University, Chengdu, Sichuan 610031, China \\
  $^{\ast}$ Corresponding author 1 Ping Zhu: zhup@hust.edu.cn \\
  $^{\ast\ast}$ Corresponding author 2 Yawei Hou: arvayhou@ustc.edu.cn \\
}

\date{\today}
\textbf{Abstract}\\
In the presence of energetic particles (EPs) from auxiliary heating and burning
plasmas, fishbone instability and Alfv\'en modes can be excited and their
transition can take place in certain overlapping regimes.
Using the hybrid kinetic-magnetohydrodynamic model in the NIMROD
code, we have identified such a transition between the fishbone
instability and the $\beta$-induced Alfv\'en Eigenmode (BAE)
for the NBI heated plasmas on HL-2A. When the safety factor at
magnetic axis is well below one, typical kink-fishbone transition occurs
as the EP fraction increases. When $q_0$ is raised to approaching one, the
fishbone mode is replaced with BAE for sufficient amount of EPs.
When $q_0$ is slightly above one, the toroidicity-induced Alfv\'en
eigenmode (TAE) dominates at lower EP pressure, whereas BAE dominates
at higher EP pressure.


\section{\label{Introduction}Introduction}

Energetic particles (EPs) produced from auxiliary heating and
burning plasmas are known to have strong influence on the
internal $1/1$ kink\cite{Shafranov1970Hydromagnetic,
  Rosenbluth1973Nonlinear, Bussac1975Internal}
and Alfv\'en eigenmodes\cite{Fasoli2007} in tokamaks. In particular,
the fishbone modes have been observed widely on tokamaks, such as
PDX\cite{mcguire1983study},
DIIID\cite{Heidbrink1990},
JET\cite{Nave1991},
HL-2A\cite{Chen2010}, and
EAST\cite{Xu2015},
in the presence of Neutral Beam Injection (NBI),
Electron Cyclotron Resonant Heating (ECRH), or
Ion Cyclotron Resonant Heating (ICRH),
which are believed to be driven by
the resonance between the
trapped EPs and the internal kink modes,
along with the diamagnetic dissipations
from the EPs\cite{chen1984excitation, coppi1986theoretical}.
On the other hand, various Alfv\'en eigenmodes (AEs), either
in the spectral gap or on the continuum, can be
excited by the EPs. Whereas the EP driven
mechanisms for each of the individual MHD and AE
instabilities have been intensively studied\cite{Rosenbluth1975Excitation,
Tsang1981Destabilitization, Fu1989Excitation, Fu1989Stability,
Chen2016Physics}, their overlapping regimes or transition conditions
have been less clear.\par
HL-2A is a medium-sized tokamak where, with ECRH and NBI
heatings, a number of EP driven instabilities, such
as ion-fishbone\cite{Chen2018Stabilization},
e-fishbone\cite{YuLM2013frequency},
TAE\cite{YuLM2018toroidal}, and
BAE\cite{Shi2019Beta},
have been observed. Based on the HL-2A configuration,
we have recently found an overlapping regime
and condition for the transition from fishbone
to BAE instabilities in the presence of EPs, using
the hybrid kinetic-MHD (HK-MHD) model implemented in the
NIMROD code\cite{sovinec2004nonlinear, Kim2008Impact}. With
continuous variation of the safety factor
profile and the EP $\beta$ fraction, the
dispersion relation and mode structure of the
dominant EP-driven instability alternate
between the characteristics of the fishbone
and the Alfv\'en modes. The overlapping or
adjacency for the regimes of
these two distinctively different EP
modes may not be always guaranteed,
and our findings may help their
identification and interpretation in experiments.\par
The rest of paper is organized as follows. The
simulation model is reviewed briefly in
section 2. The simulation set-up is described
in section 3, which is followed by the report
on the main results in section 4. Finally,
we conclude with a summary and discussion in
section 5. \par

\section{Simulation model}
The HK-MHD equations implemented in the \nimrod\ code
including EP effects are as follows\cite{Kim2008Impact}
\begin{equation}
  \frac{\partial\rho}{\partial t} + \nabla\cdot(\rho\mbV) = 0
\end{equation}
\begin{equation}
  \rho \left(\frac{\partial\mbV}{\partial t}+\mbV\cdot
  \nabla\mbV\right) = \mbJ\times\mbB-\nabla p_b-\nabla\cdot\bm P_h
\end{equation}
\begin{equation}
  \frac{n}{\gamma-1} \left(\frac{\partial T}{\partial t}
  +\mbV\cdot\nabla T\right) = 0
\end{equation}
\begin{equation}
  \frac{\partial\mbB}{\partial t}=-\nabla\times\mbE
\end{equation}
\begin{equation}
  \mu_0\mbJ=\nabla\times\mbB
\end{equation}
\begin{equation}
  \mbE=-\mbV\times\mbB
\end{equation}
where $\rho$, $\mbV$, $\mbJ$, $\mbB$, $p_b$, and  $\mbE$ are the
mass density, center of mass
velocity, current density, magnetic field,
background pressure, and  electric field respectively.
In the limits of $n_h\ll n_b$, $\beta_h\sim \beta_b$, where $n_b$ ($n_h$) is
the background plasma (energetic particle) number density,
$\beta_b$ ($\beta_h$) is the ratio of background plasma (energetic particle)
pressure to magnetic pressure, and the kinetic effect of EP is coupled
into MHD equations by adding EP pressure tensor
$\bm P_h$ to the momentum equation.
The drift-kinetic equation is solved to determine the EP distribution and
the pressure tensor $\bm P_h$\cite{Kim2008Impact}.
\section{Simulation setup}
\begin{figure}[ht]
  \centering
  \includegraphics[width=7.2cm]{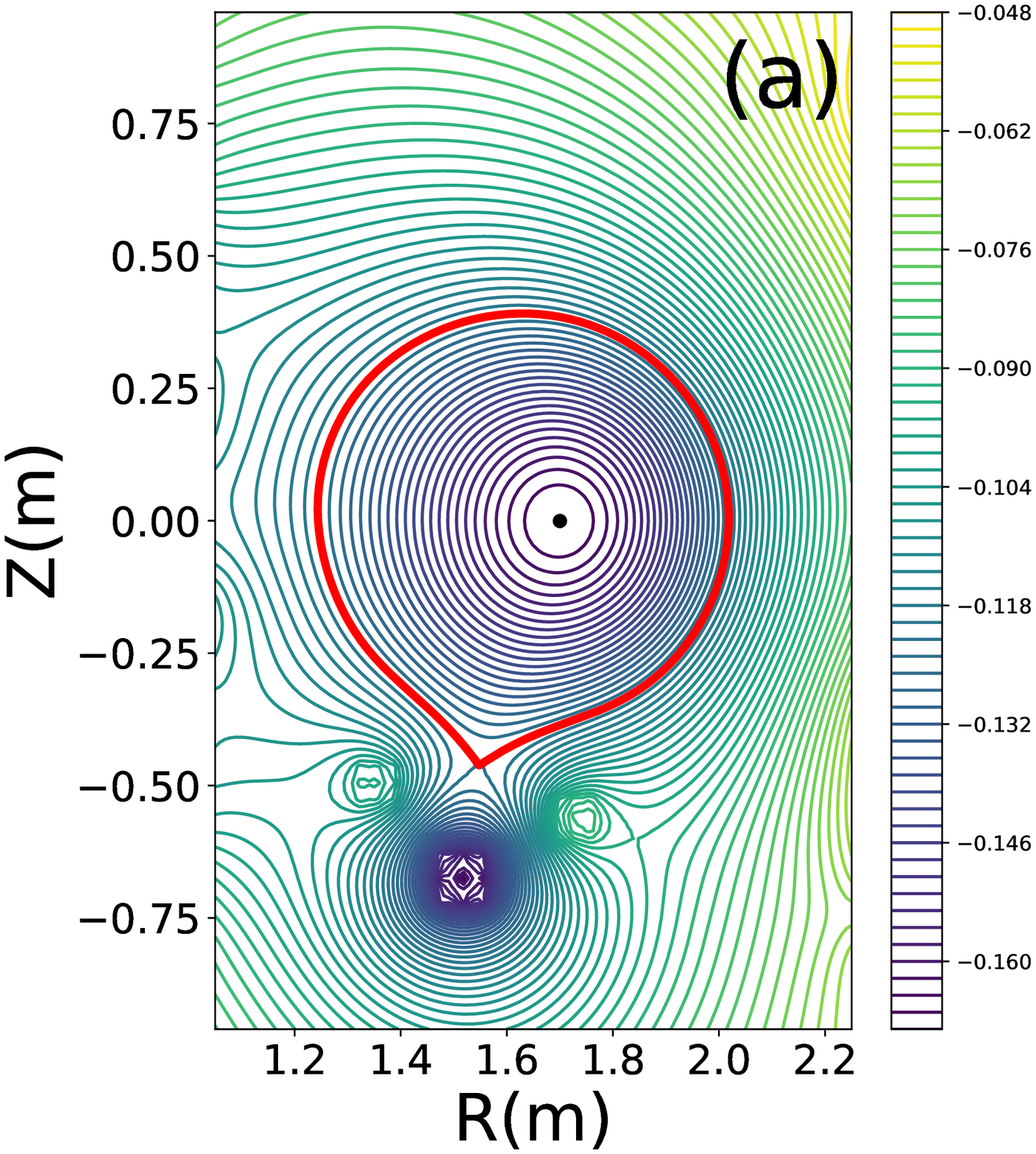}
  \includegraphics[width=6.8cm]{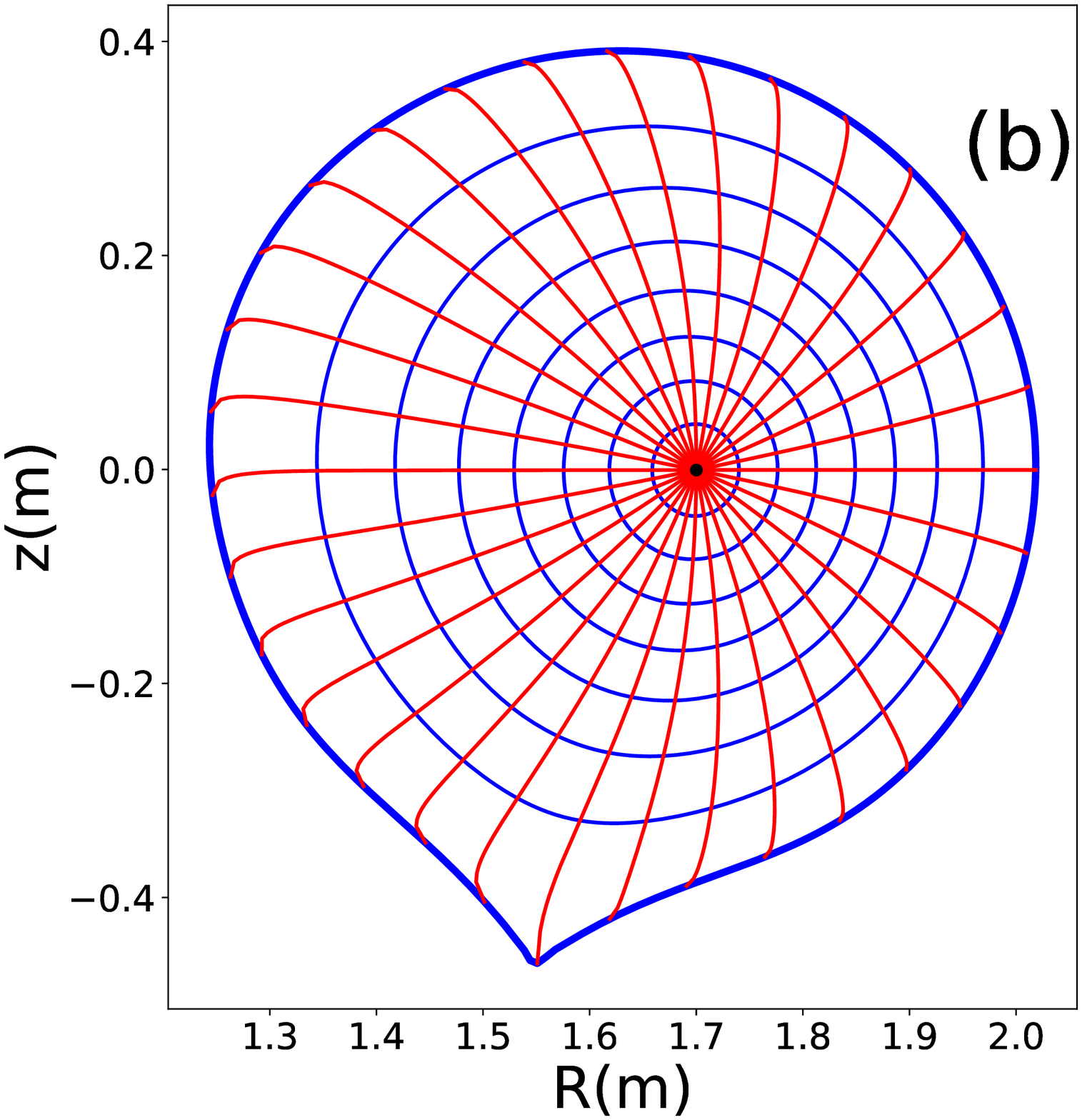}
  \caption{\label{fig:psicon} (a) Contour plot of equilibrium poloidal
    flux in ($R, Z$) coordinate. The red curve
    represents the last closed flux surface (LCFS).
    (b) The mesh grid of flux coordinates used in the Alfv\'en
    continuum calculation. Uniform poloidal flux and equal poloidal arc
    length are used in this flux coordinate system. The black point locates
    at the magnetic axis. The blue lines represent the constant
    poloidal fluxes, and the red lines the constant poloidal angles.
    The mesh grids used in the actual
    calculation are much finer than the above diagram.}
\end{figure}

\begin{figure}[ht]
  \centering
  \includegraphics[width=8.0cm]{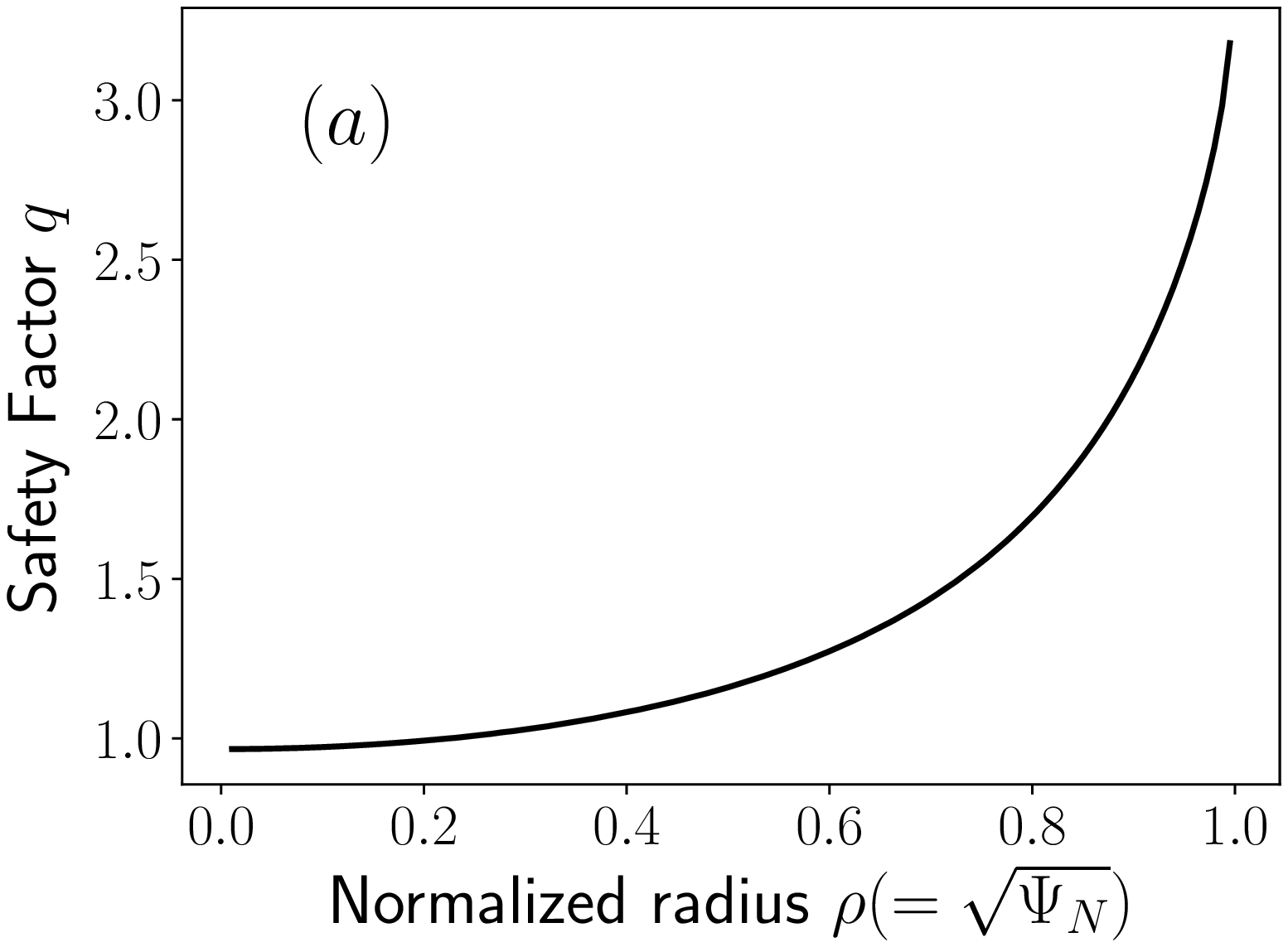}
  \includegraphics[width=8.0cm]{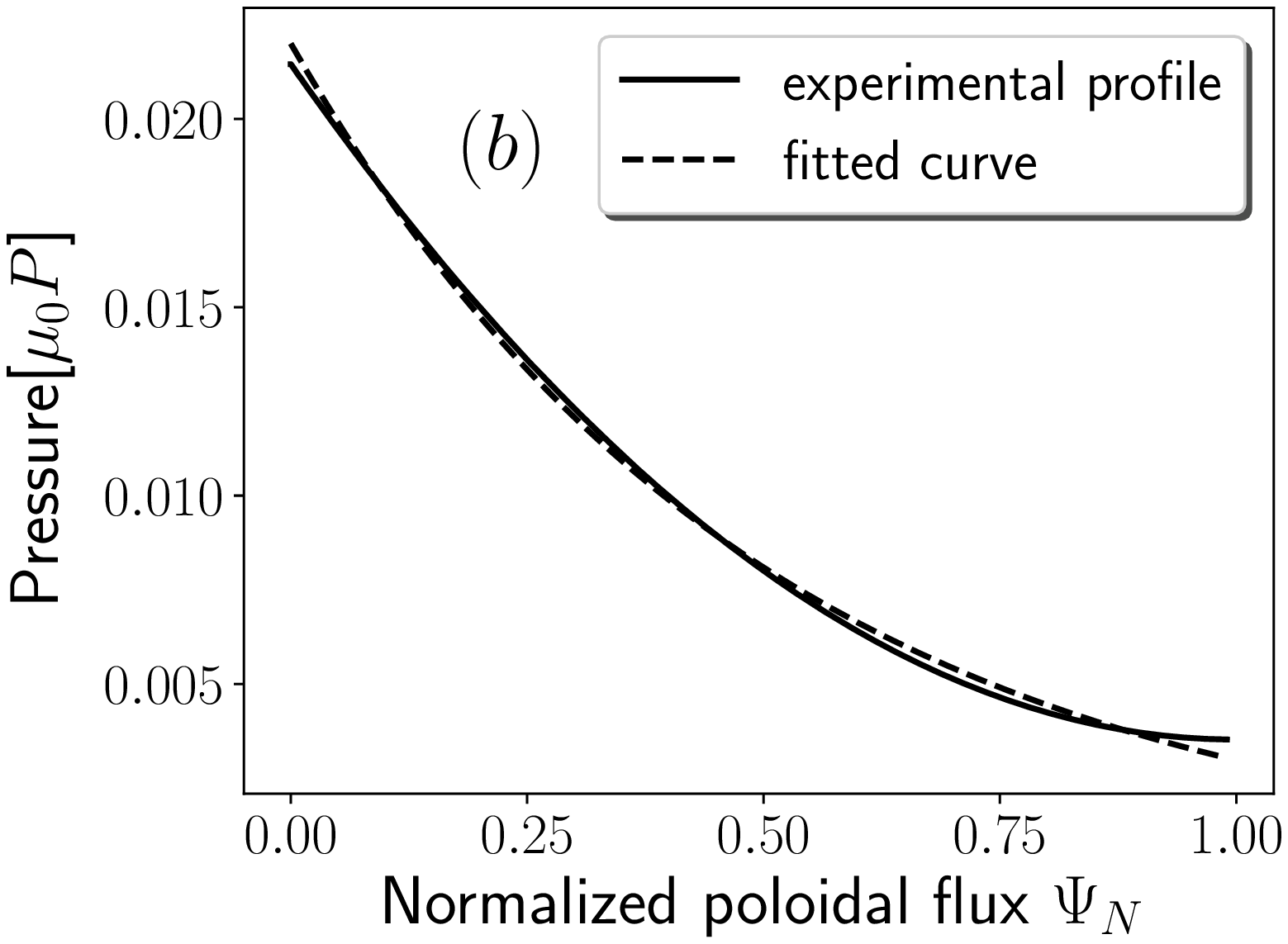}
  \caption{\label{fig:qprof}(a) $q-$profile and
  (b) pressure profiles in HL-2A discharge \#16074.}
\end{figure}

The equilibrium is constructed from HL-2A discharge
\#16074 using \efit\cite{DengWei2014investigation}.
The equilibrium flux surfaces and the mesh grid in magnetic flux coordinates
are shown in FIG \ref{fig:psicon}, where the last
closed flux surface (LCFS) is up-down asymmetric, and the flux
surfaces are close to circles  in the core region.
As can be seen in FIG \ref{fig:qprof}(a),
the $q$ profile is rather flat about one in the core region
$0<\sqrt{\psi/\psi_0}<0.2$,
where $\psi$ is the poloidal magnetic flux,
and $\psi_0$ is the total poloidal magnetic flux within the LCFS.
The slowing-down distribution function is used for the EPs,
\begin{equation}
  f_0 = \frac{P_0\exp \left(\frac{P_{\zeta}}{\psi_n}\right)}{
    \varepsilon^{3/2}+\varepsilon_c^{3/2}}
\end{equation}
where $P_0$ is the normalization constant,
$P_{\zeta}=\tg\rho_{\parallel}-\psi_p$ is the canonical toroidal momentum,
$\tg=RB_{\phi}$, $\rho_{\parallel}=mv_{\parallel}/qB$, $\psi_p$ is
the poloidal flux, $\psi_n = c\psi_0$ ,
$\psi_0$ is the total flux and the parameter $c$
is used  to match the spatial
profile of the equilibrium, $\varepsilon$ is the particle energy,
and $\varepsilon_c$ is the critical slowing
down energy\cite{Goldston00}
\begin{equation}\label{eq:critdef}
  \varepsilon_c = \left(\frac{3}{4}\right)^{2/3}\left(\frac{\pi m_i}{m_e}
  \right)^{1/3} T_e
\end{equation}
with $m_i$ being the ion mass, $m_e$ the electron mass, and $T_e$ the electron
temperature. When $\varepsilon>\varepsilon_c$, the slowing down of
beam ions is mainly due to the collisions with background electrons,
and the collisions with background ions is dominant when
$\varepsilon<\varepsilon_c$. As the beam ions slow down, they give up
their energy increasingly to the background ions, rather than to the
background electrons. The slowing down distribution models
the process which is dominated by the collisions
between beam ions and background ions.\par
As shown in FIG \ref{fig:qprof}(b), the energetic particles are loaded into
the physical space following  the profile $p=p_0\exp(-c\psi/\psi_0)$,
where $p_0$ is pressure at magnetic axis, $c=0.5$.
Parameters $p_0$ and $c$ are generated
from the fitting to the original pressure profile
from experiment. Other main parameters are set up as
follows\cite{Zhang2014Long}. The major radius $R=1.65\mm$, and the minor
radius $a=0.40\mm$ for the LCFS, the toroidal magnetic field
$B_0=1.37\mT$, the number density $n=2.44\times 10^{19}\mm^{-3}$,
the initial energy of beam ions is $\varepsilon_m=40\mkeV$,
the background electron temperature $T_e=1\mkeV$, and according
to equation \eqref{eq:critdef}, the critical energy is
$\varepsilon_c=14.8\mkeV$.
\section{Simulation results}
\subsection{Effects of $q_0$ in absence of energetic particles}
In absence of energetic particles, we find in
\nimrod\ calculations that the $(m,n)=(1,1)$ mode is the most unstable,
where $n$ and $m$ are the toroidal and the poloidal mode numbers respectively.
\begin{figure}[ht]
  \centering
  \includegraphics[width=8.0cm]{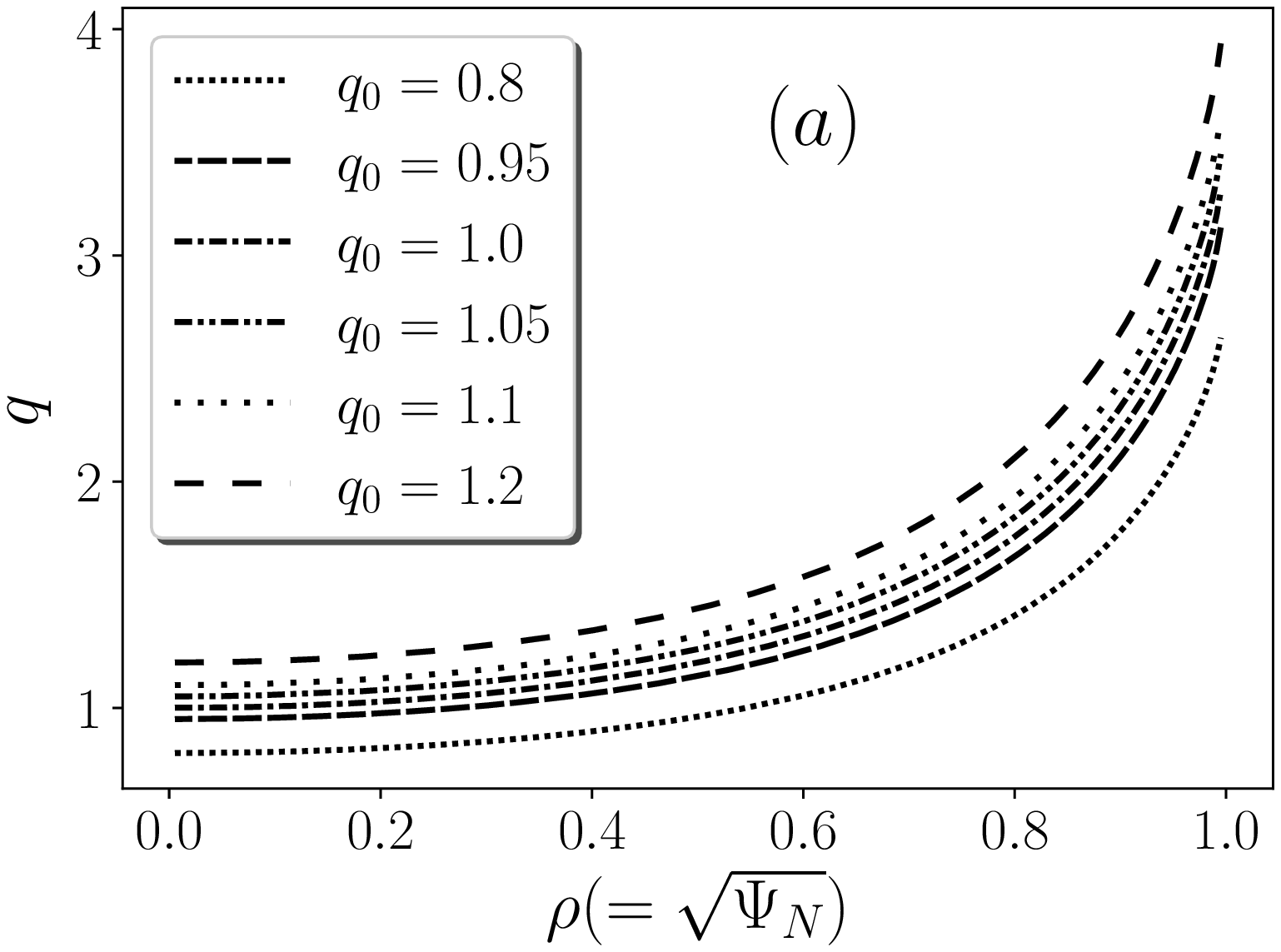}
  \includegraphics[width=8.0cm]{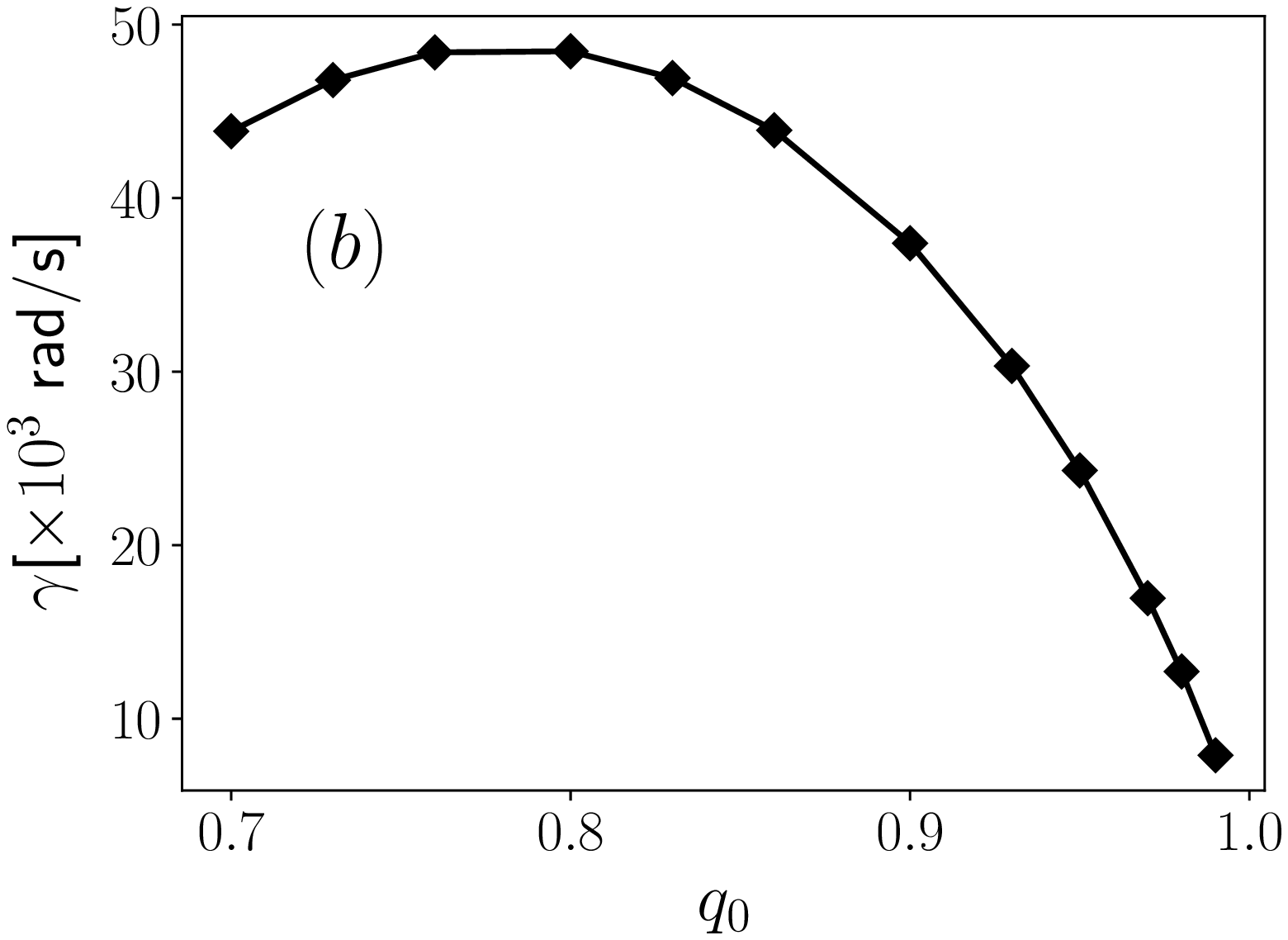}
  \caption{\label{fig:q0eff}(a) $q$-profiles with different
    $q_0$ and (b) the growth rates dependence on $q_0$
    for $(m,n)=(1,1)$ modes.}
\end{figure}
Only the safety factor $q_0$ at magnetic axis is varied, so that
the $q$-profile shifts up or down entirely without changing its
shape (FIG. \ref{fig:q0eff}a). The linear growth rate of the
$(1,1)$ mode from NIMROD calculations increases first before decreases,
and approaches  $0$ as $q_0$ approaches to unity. The growth rate
of $(1,1)$ modes reaches its maximum at $q_0=0.77$. \par
Such a dependence of growth rate on $q_0$ may be understood from
previous theory\cite{Bussac1975Internal}.
For a monotonic parabolic $q$-profile with $q_0$ below 1,
the distance $|1-q_0|$ can be seen as a
measure of free energy within the $q=1$ surface
from equation
\begin{equation}\label{eq:bussac1975}
  \gamma = -\frac{r_0V_A\pi\delta W_T}{\sqrt{3R_0}R_0 q'}
\end{equation}
\begin{equation}\label{eq:delWT}
  \delta W_T \sim 3\nu\Delta q \left[\frac{13}{48(\nu+4)} -
    \beta_p^2\int_{r_0}^{r^2}
    \frac{\dif r}{r_0}\left(\frac{r}{r_0}\right)^{\nu-5}\right]
\end{equation}
where $\Delta q = 1-q_0$, $1-q(r)\sim \Delta q[1-(r/r_0)^\nu]$,
$q(r_0)=1$, $V_A$ is Alfv\'en velocity, $R_0$ is the major radius at magnetic
axis, and $\beta_p$ = $[2\mu_0/B_p^2(r_0)]\int_0^{r_0}(r/r_0)^2(-\dif p/\dif r)$
is poloidal beta.  It follows that the free energy
decreases with $q_0$ ($q_0<1$).
The magnetic shear $q'=\dif q/\dif r$ and the toroidal potential energy
$\delta W_T$ (free energy) can stabilize and destabilize the mode respectively.
As $q_0$ increases, both the magnetic shear $q'$ and the toroidal potential
energy $\delta W_T$ decrease. At first, the decreasing of stabilizing
effect from the  magnetic shear $q'$ is dominant, so
the growth rate increases; as $q_0$ increases further,
the reduction of toroidal potential energy $\delta W_T$ becomes
dominant, so the growth rate starts to decrease. \par
\begin{figure}[ht]
  \centering
  \begin{overpic}[scale=0.25]{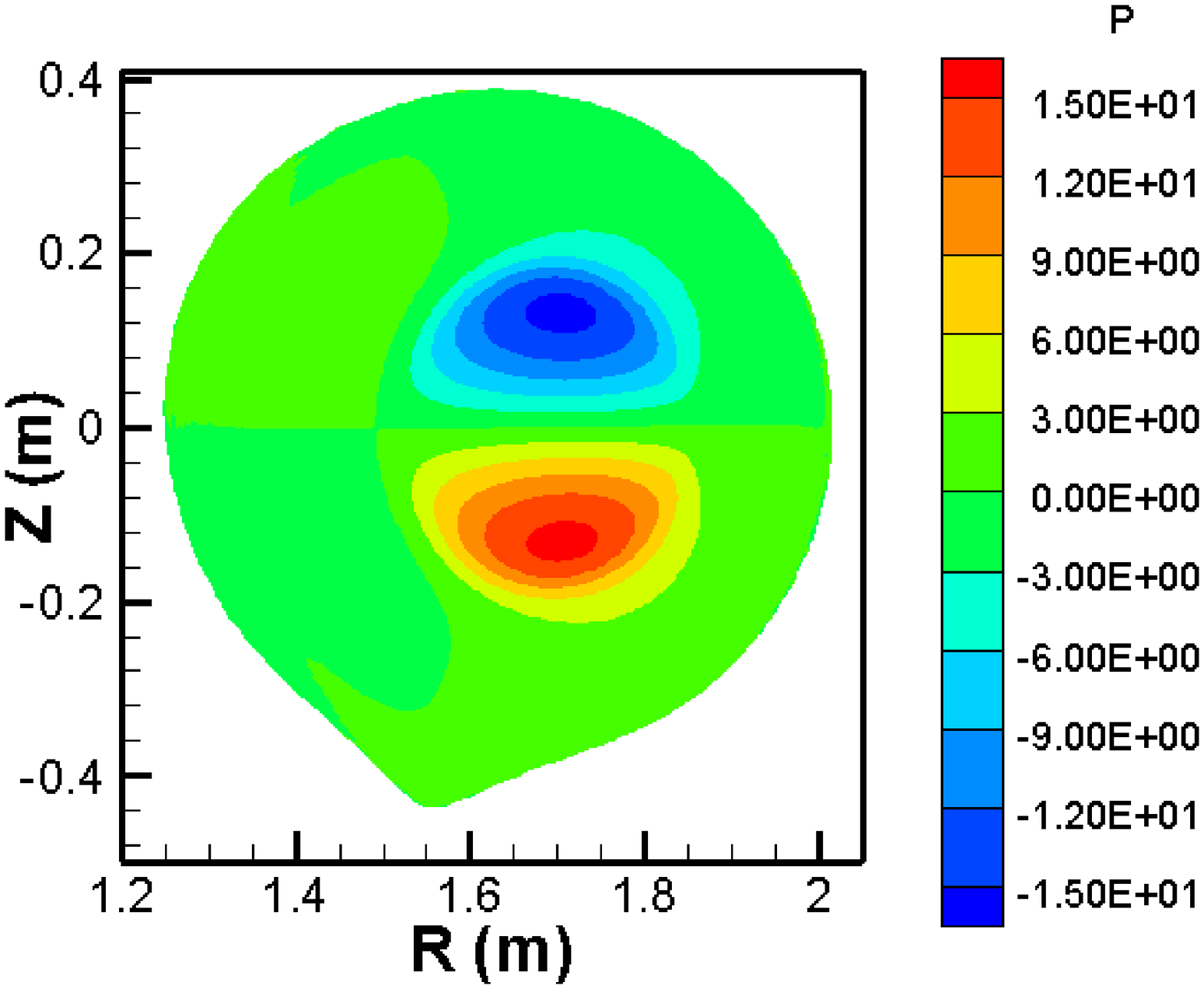}
    \put(13, 65){$(a)$}
  \end{overpic}
  \begin{overpic}[scale=0.25]{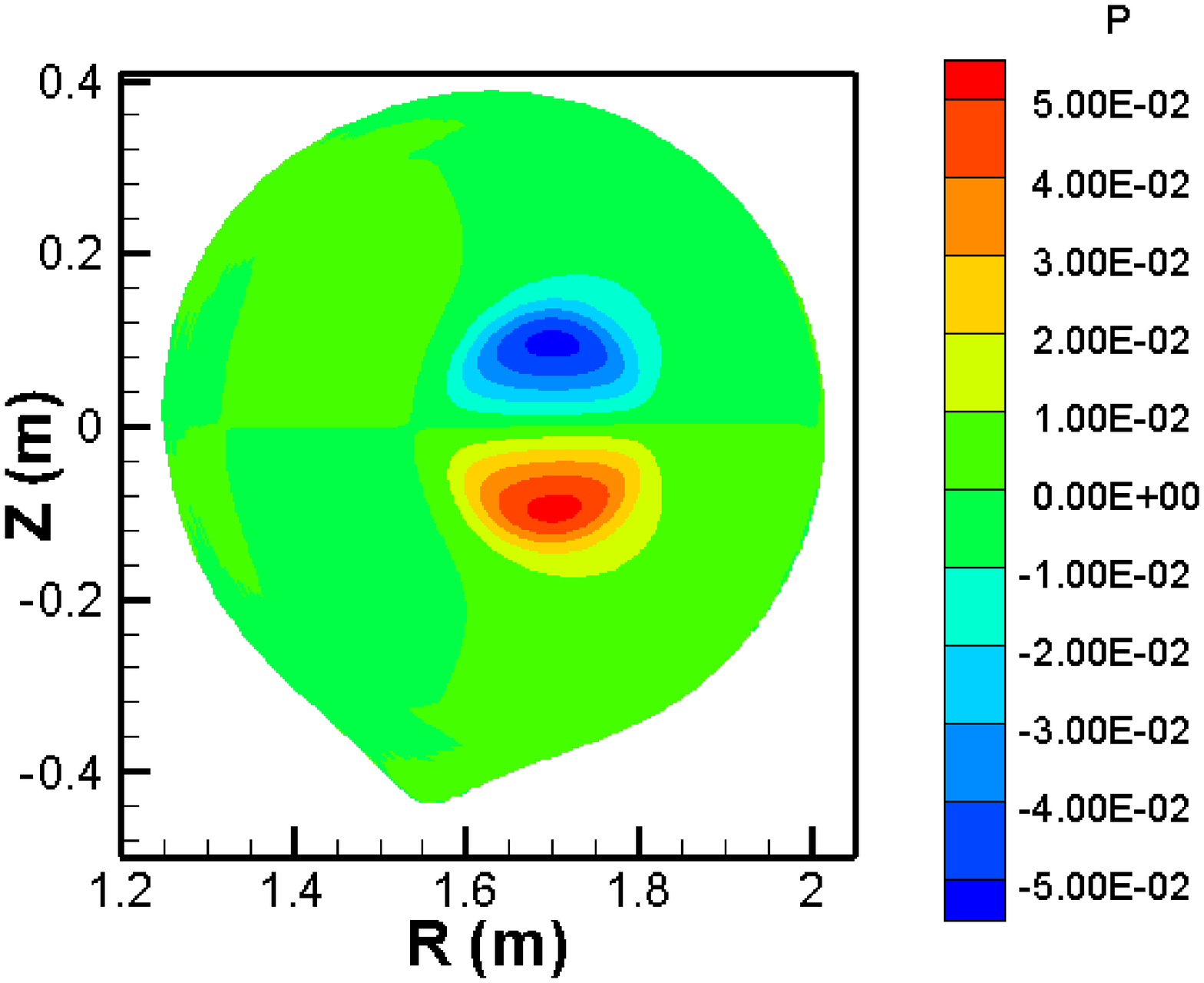}
    \put(13, 65){$(b)$}
  \end{overpic}
  \begin{overpic}[scale=0.25]{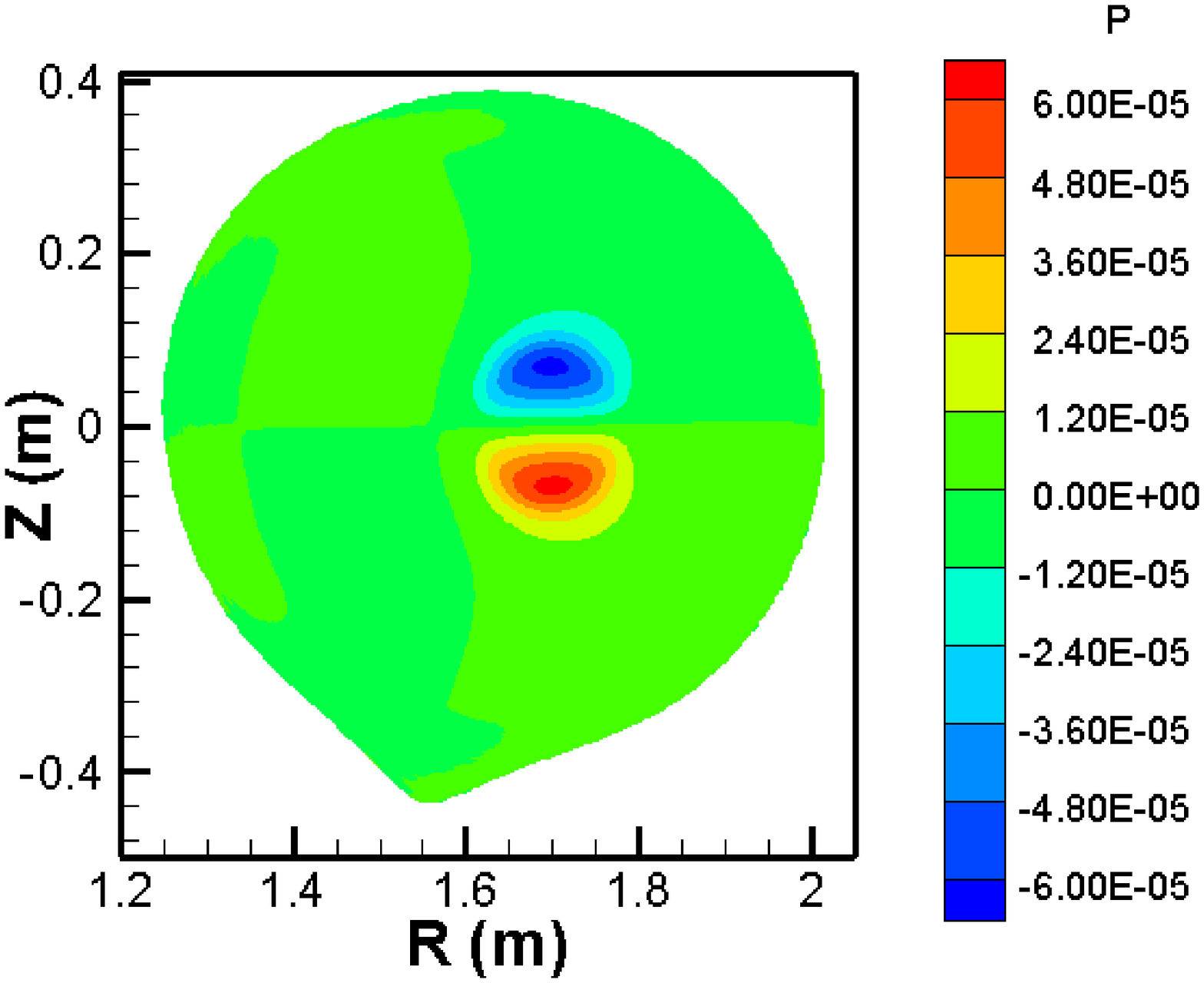}
    \put(13, 65){$(c)$}
  \end{overpic}
  \caption{\label{fig:n1q0s} Contour plots of pressure perturbation
    for different $q_0$: (a) $q_0=0.80$, (b) $q_0=0.90$, and (c) $q_0=0.95$.}
\end{figure}
The contour plots of the plasma pressure perturbation show that the
$(1, 1)$ mode structure shrinks in size and becomes more
localized in the core region as $q_0$ approaches to $1$,
which further confirms the theory prediction in Equations
\eqref{eq:bussac1975} and \eqref{eq:delWT}. \par
\subsection{Effects of $q_0$ in the presence of energetic particles}
\begin{figure}[ht]
  \centering
  \includegraphics[width=8.0cm]{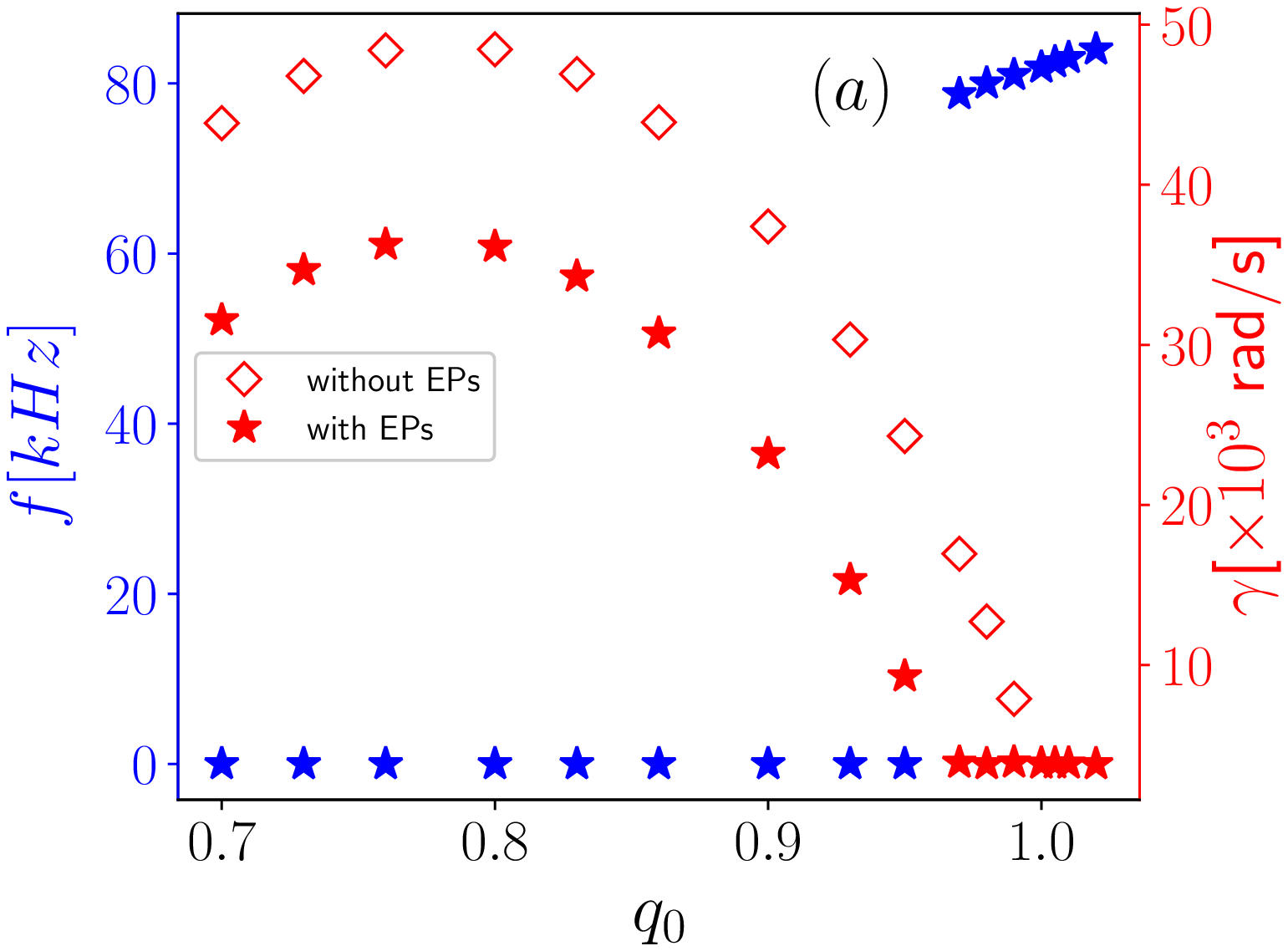}
  \includegraphics[width=8.0cm]{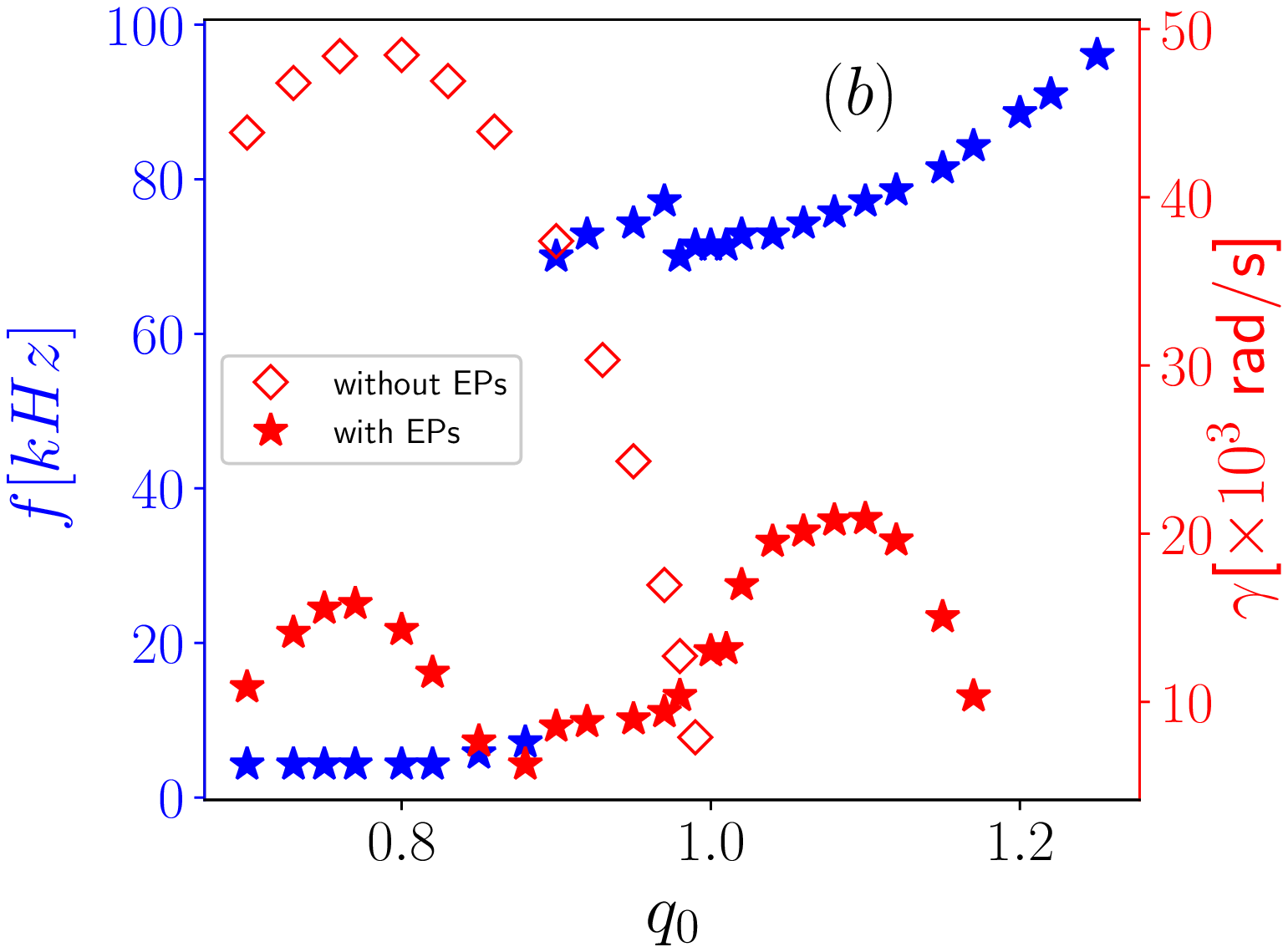}
  \caption{\label{fig:fgbhsn11} The frequency and growth rate dependence on
    $q_0$  for $(1,1)$ modes. (a) $\beta_h/\beta_0=0.25$ and
    (b) $\beta_h/\beta_0= 0.5$.}
\end{figure}
\begin{figure}[ht]
  \centering
  \begin{overpic}[scale=0.33]{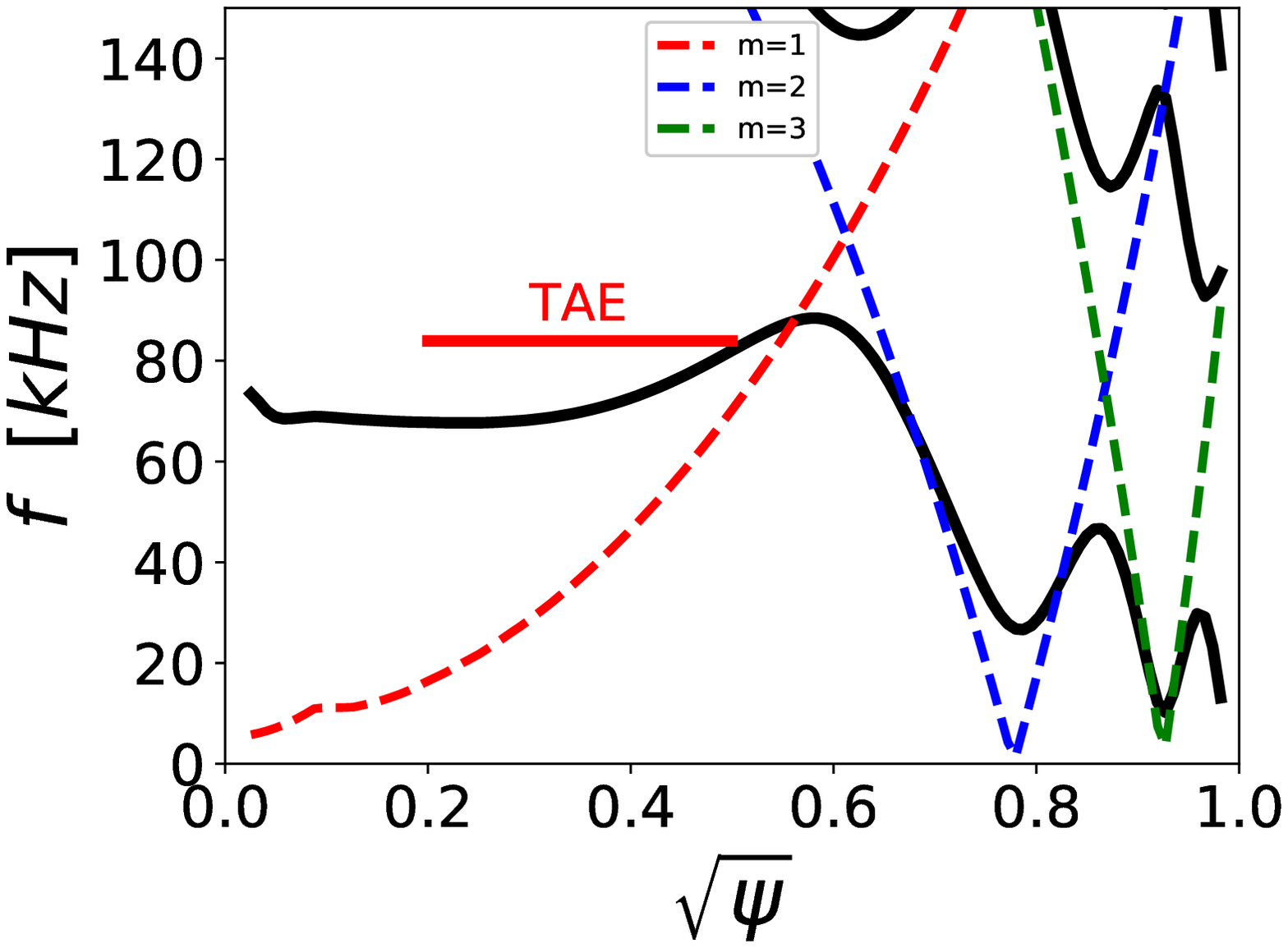}
    \put(23, 65){$(a)$}
  \end{overpic}
  \begin{overpic}[scale=0.33]{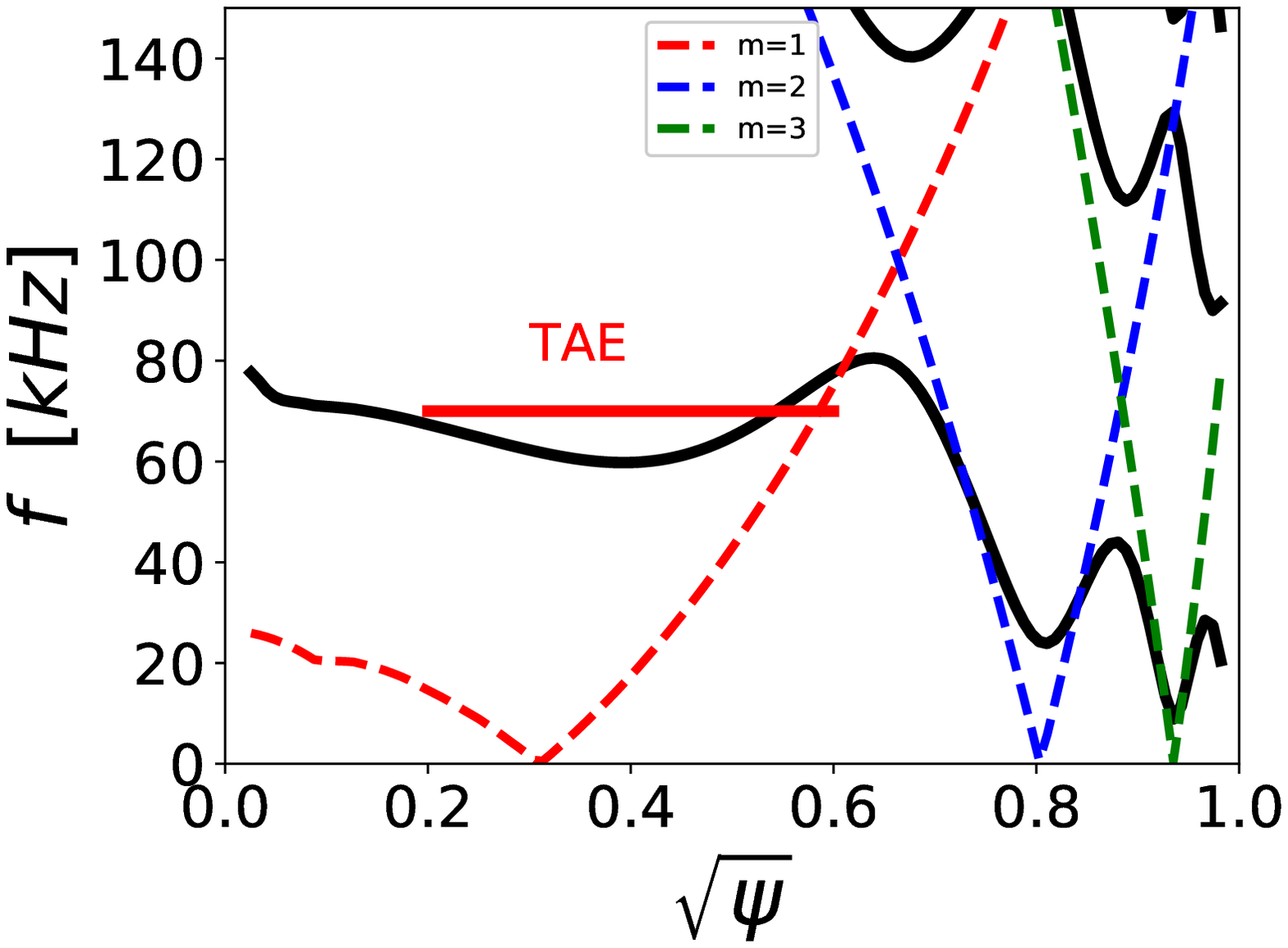}
    \put(23, 65){$(b)$}
  \end{overpic}
  \begin{overpic}[scale=0.33]{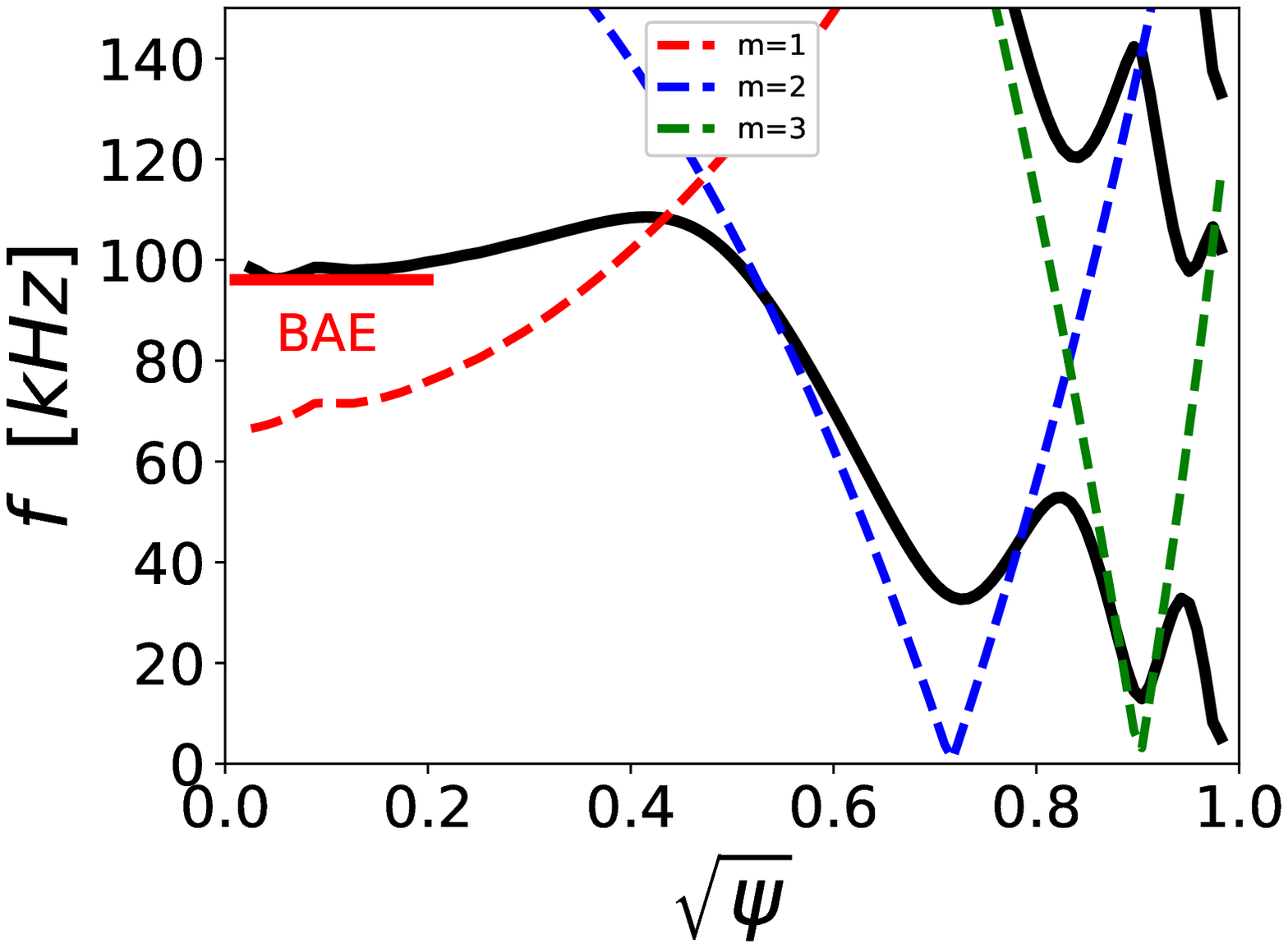}
    \put(23, 65){$(c)$}
  \end{overpic}
  \caption{\label{fig:acqbh} Alfv\'en continua with toroidal mode
    number $n = 1$ calculated based on the slow-sound approximation from
    AWEAC. Alfv\'en continua from the cylindrical geometry limit
    (dotted line) are also given. (a) $\beta_h/\beta_0=0.25$, $q_0=1.02$;
    (b) $\beta_h/\beta_0=0.5$, $q_0=0.90$; and
    (c) $\beta_h/\beta_0=0.5$, $q_0=1.25$.}
\end{figure}
In the presence of EPs with $\beta_h/\beta_0=25\%$, the growth rate of
$(1, 1)$ mode increases first and then decreases as $q_0$
increases (FIG. \ref{fig:fgbhsn11}a). Comparison with the cases without
EPs indicates that the dependence of
growth rate on $q_0$ is similar whereas the EPs have an overall
stabilizing effect. The frequency is almost constant ($3kHz$) when
$q_0<0.95$, which can be identified as that of the fishbone mode.
For $q_0>0.95$, the mode frequency jumps to another branch
around $80kHz$ and increases with $q_0$, which is considered
as a TAE from the Alfv\'enic continua in FIG. \ref{fig:acqbh}(a).
For higher EP fraction with $\beta_h/\beta_0=50\%$, the transition
from the fishbone branch to the TAE branch takes place at a lower
$q_0\simeq 0.88$, and the transition from the TAE branch to the BAE
branch happens at $q_0\simeq 0.97$, and the significantly enhanced BAE growth
rate reaches its maximum around $q_0=1.1$(FIG. \ref{fig:fgbhsn11}b).\par
\begin{figure}[ht]
  \centering
  \begin{overpic}[scale=0.33]{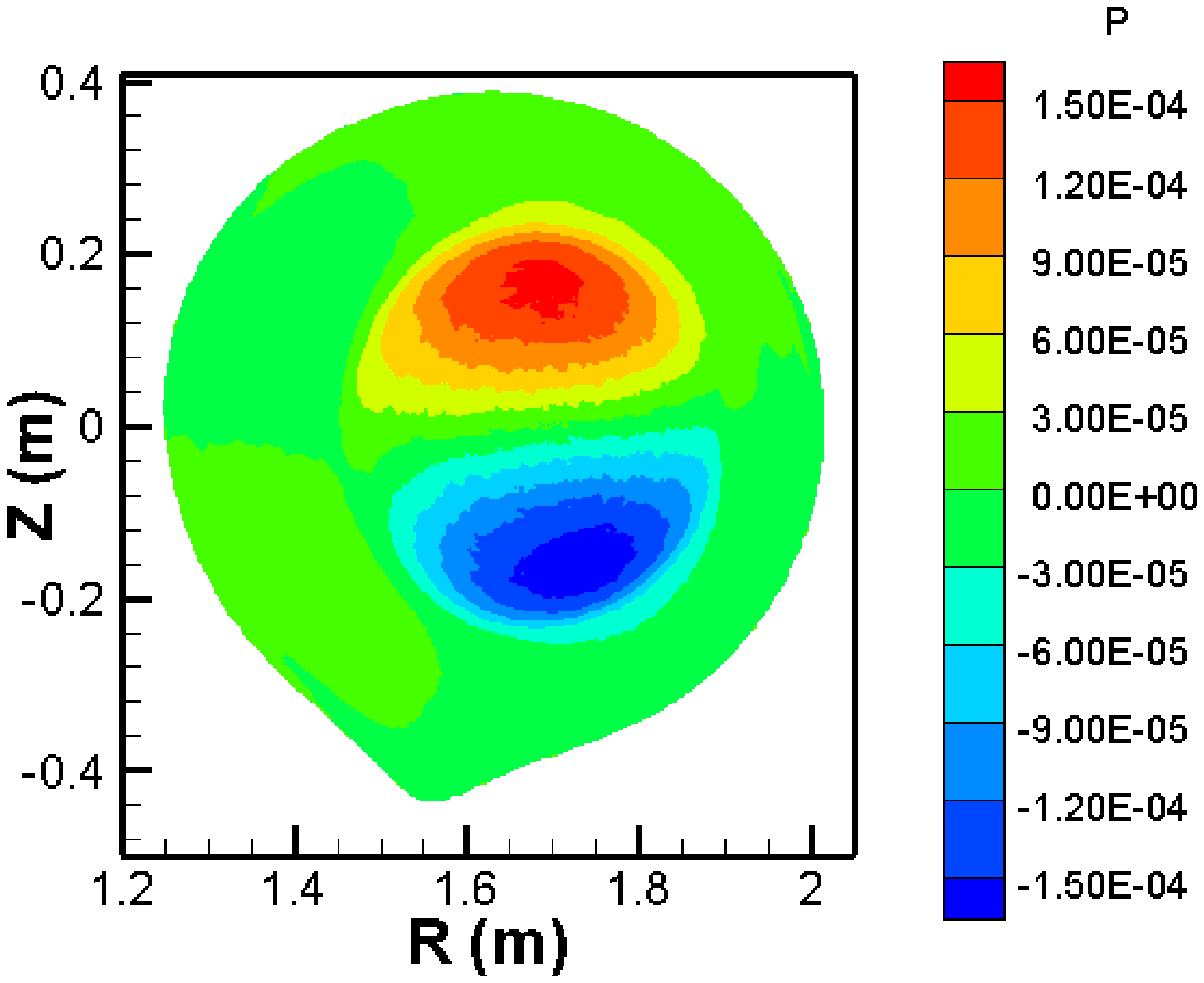}
    \put(13, 65){$(a)$}
  \end{overpic}
  \begin{overpic}[scale=0.40]{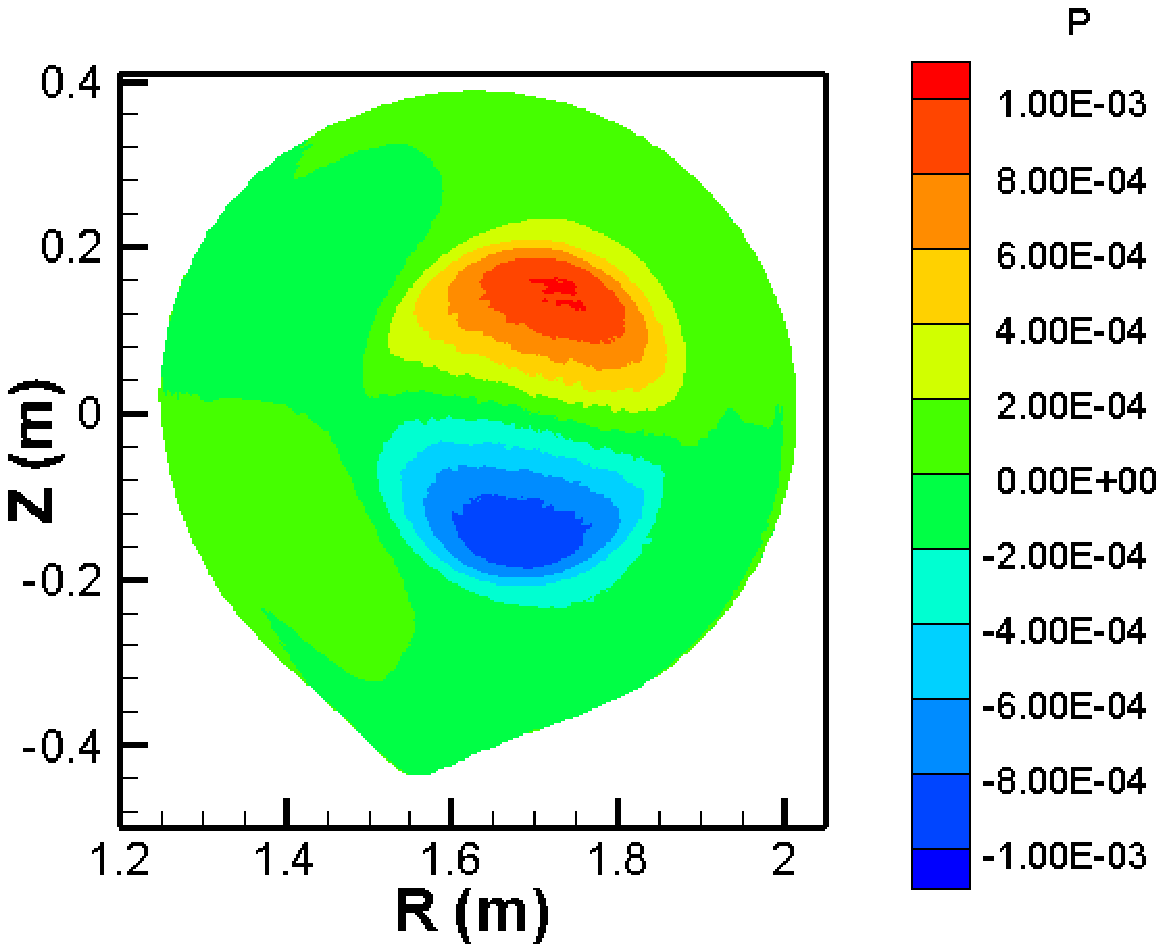}
    \put(13, 65){$(b)$}
  \end{overpic}
  \begin{overpic}[scale=0.25]{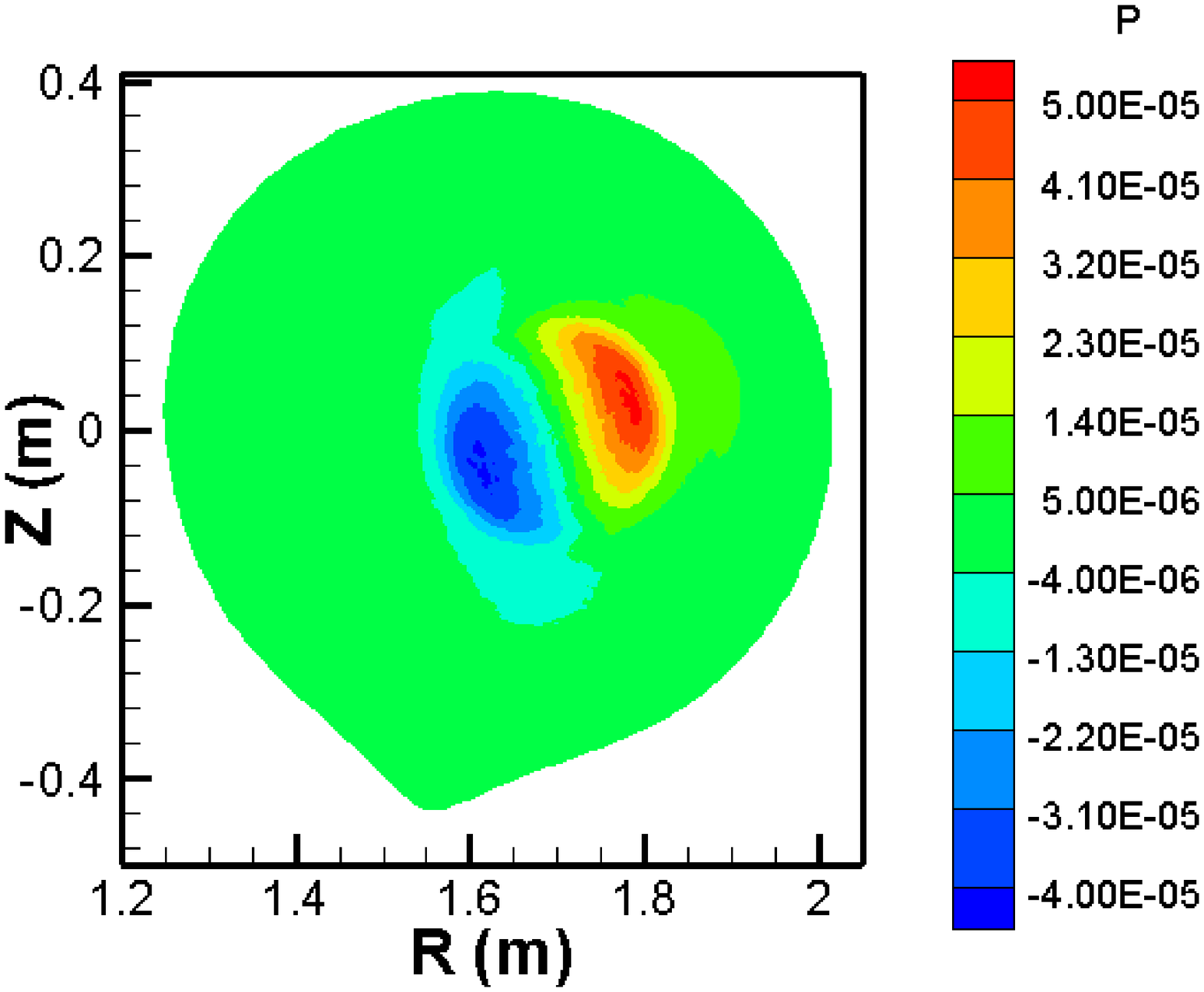}
    \put(13, 65){$(c)$}
  \end{overpic}
  \begin{overpic}[scale=0.25]{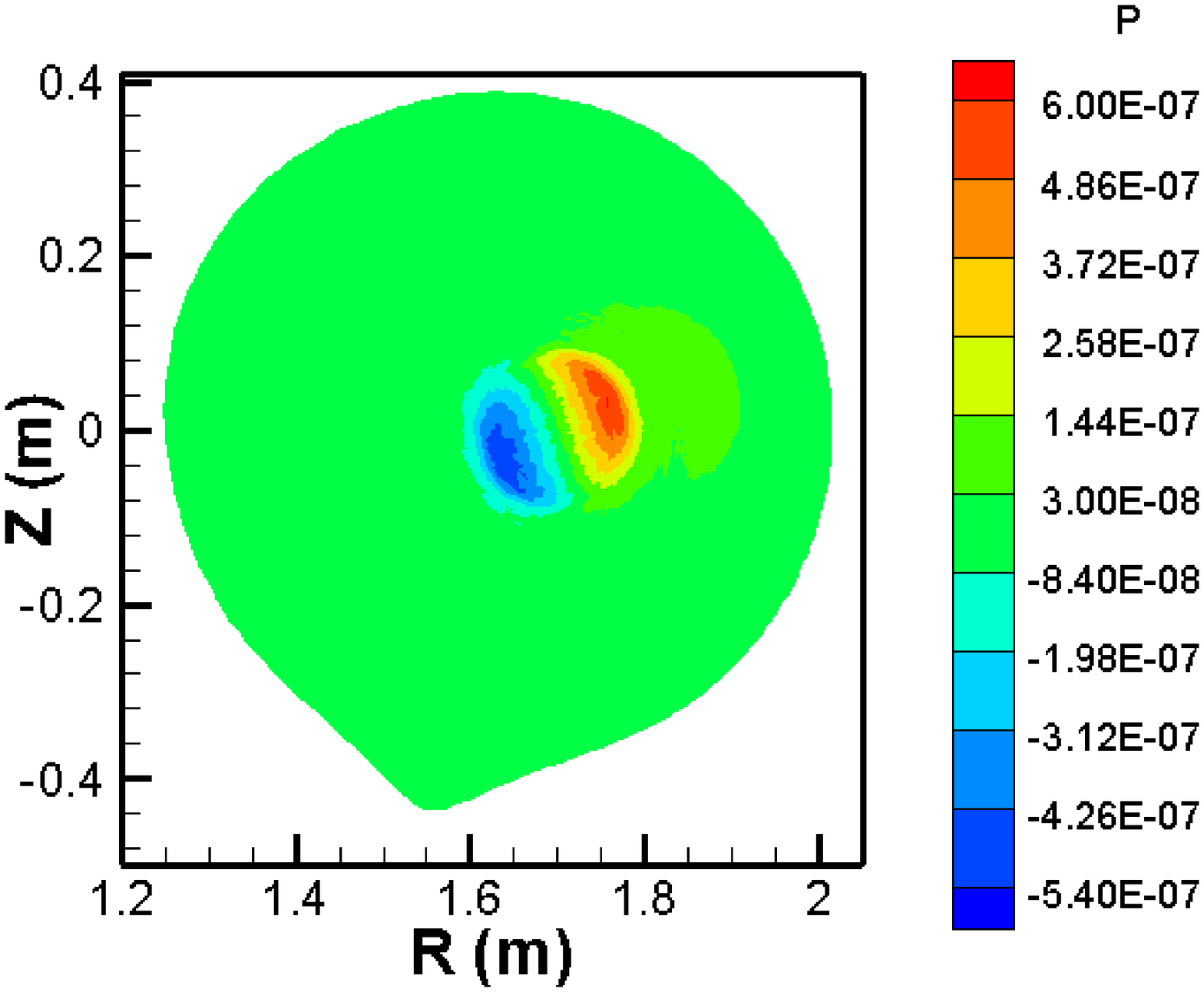}
    \put(13, 65){$(d)$}
  \end{overpic}
  \begin{overpic}[scale=0.25]{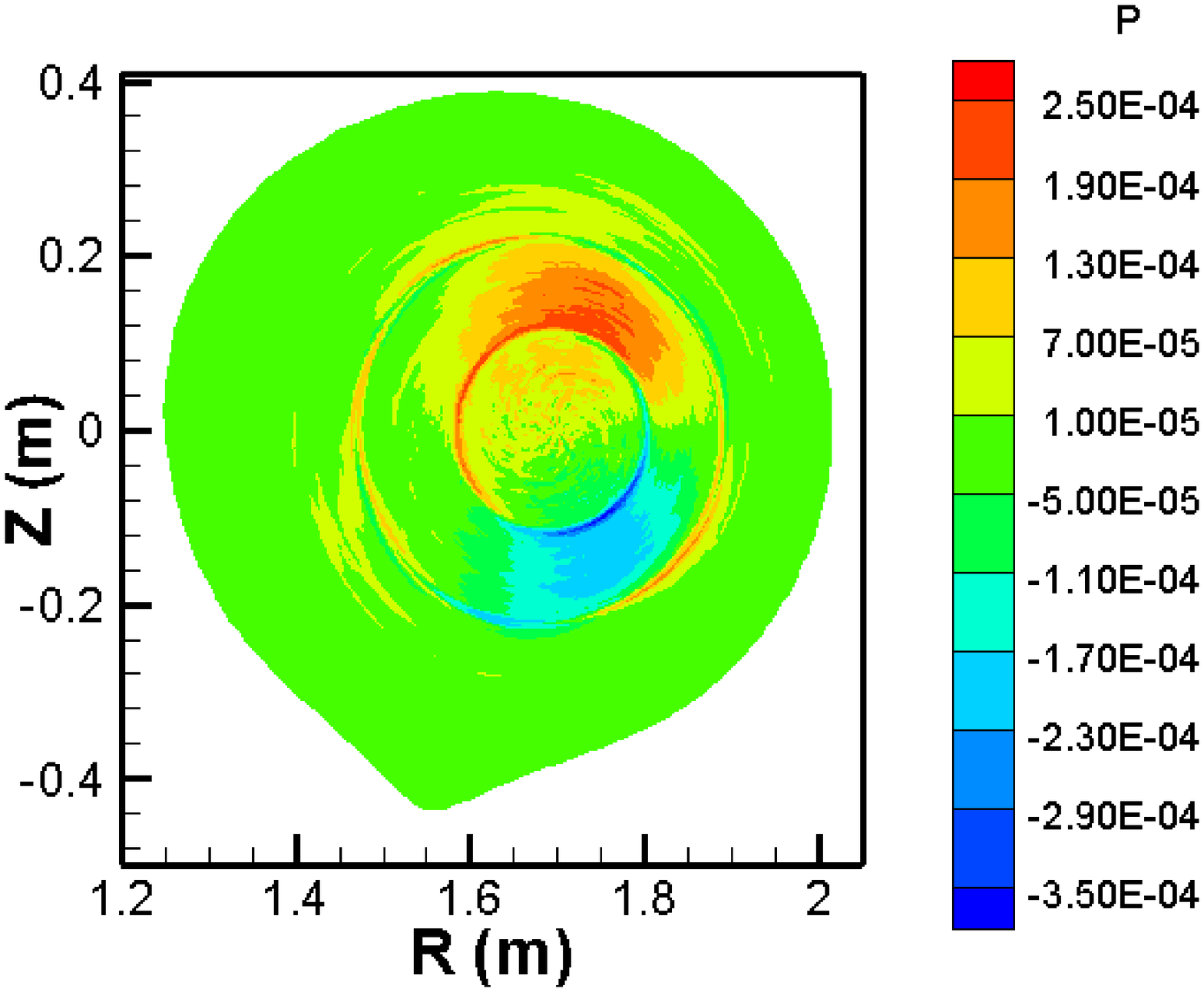}
    \put(13, 65){$(e)$}
  \end{overpic}
  \begin{overpic}[scale=0.33]{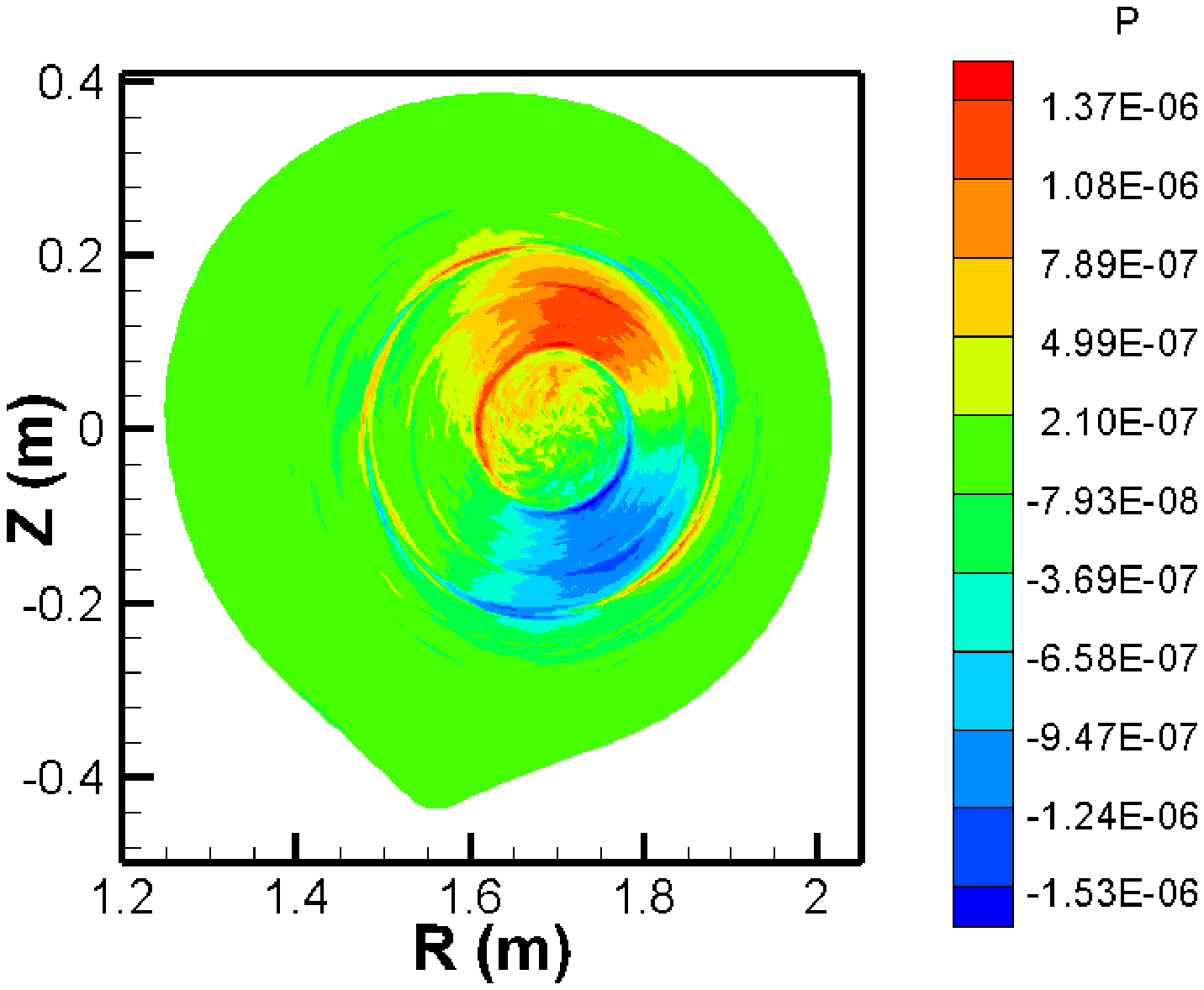}
    \put(13, 65){$(f)$}
  \end{overpic}
  \caption{\label{fig:betah25n1} Contour plots of pressure
    perturbation of $n=1$ mode in poloidal plane  with $\beta_h/\beta_0=25\%$
    and (a) $q_0=0.70$, (b) $q_0=0.77$, (c) $q_0=0.90$,
    (d) $q_0=0.95$, (e) $q_0=1.02$, and (f) $q_0=1.05$.}
\end{figure}
The dominant modes transition is also evident from the variation
of mode structure with $q_0$. For example, in the
case with $\beta_h/\beta_0=0.25$, the perturbed pressure
contour for the $n=1$ mode in the poloidal plane shows clear
$(1, 1)$ kink mode structure inside the $q=1$ surface when $q_0<0.95$, which
shrinks in size as $q_0$  increases (FIG. \ref{fig:betah25n1}a$\sim$d).
When $q_0>0.95$, the mode structure becomes qualitatively different,
which now involves the coupling between two rational surfaces, which
is characteristic of of the TAE mode (FIG. \ref{fig:betah25n1}e$\sim$f).
\subsection{Effects of energetic particles for different $q_0$}
\begin{figure}[ht]
  \centering
  \includegraphics[width=5.0cm]{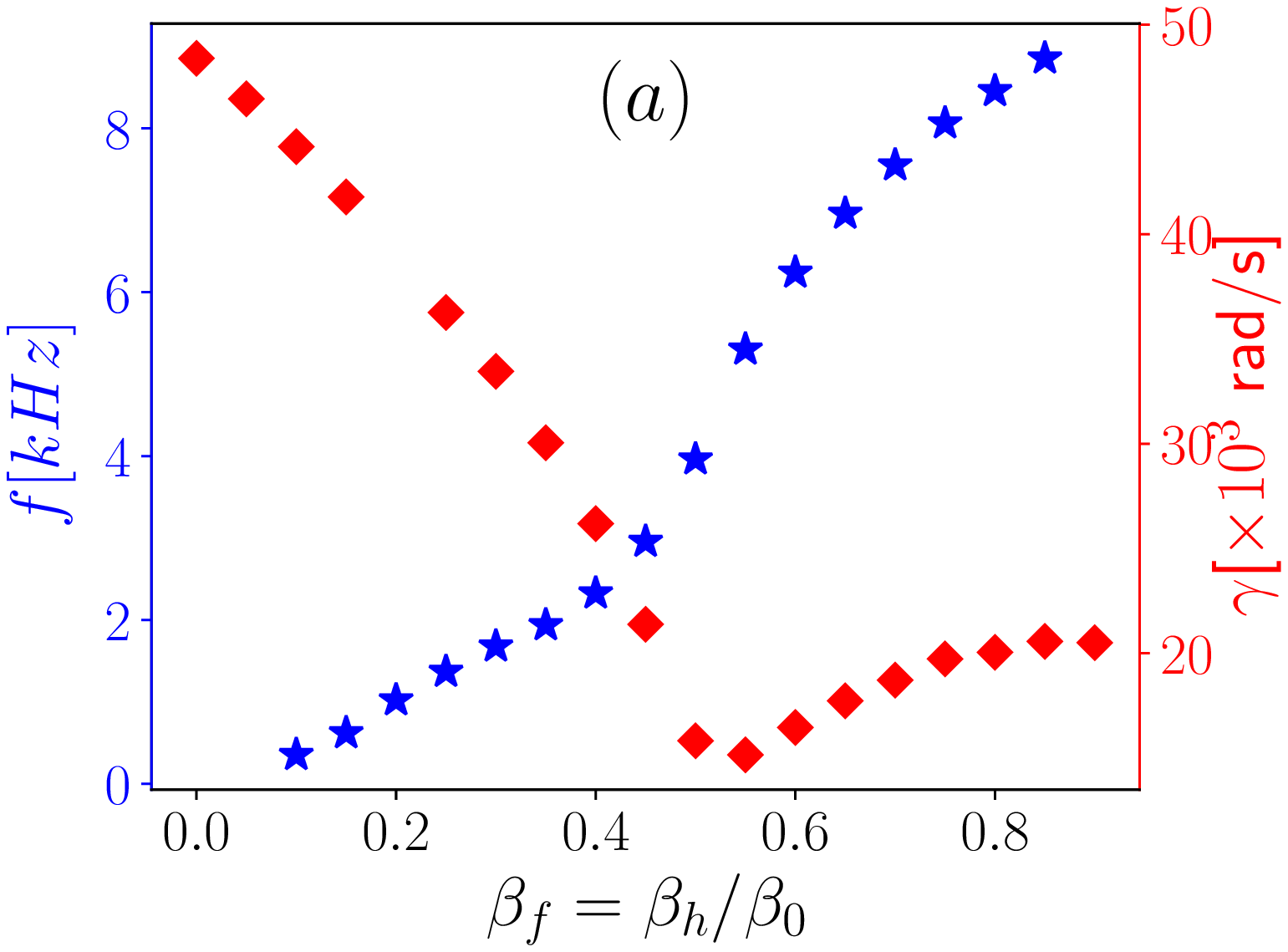}
  \includegraphics[width=5.0cm]{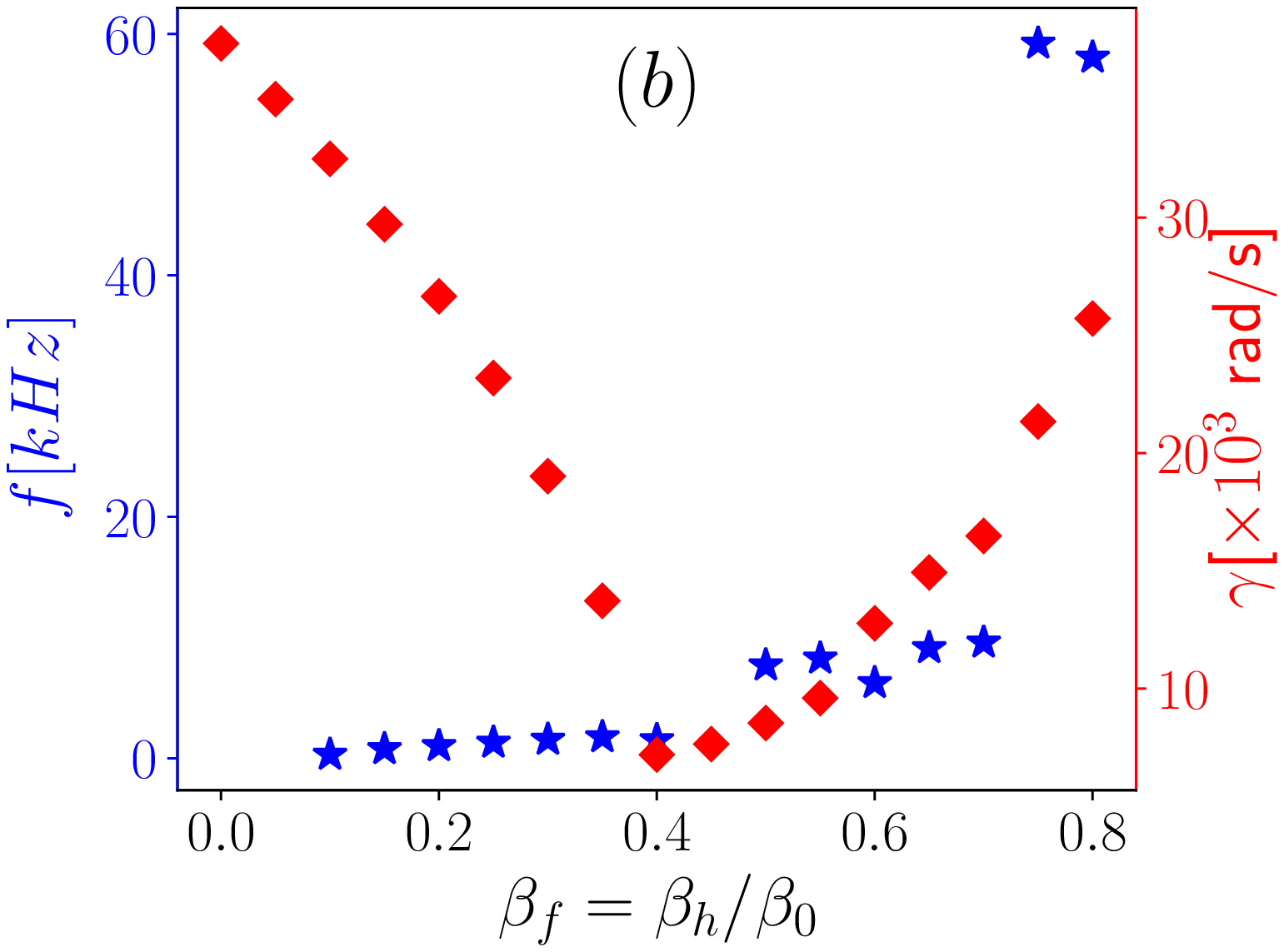}
  \includegraphics[width=5.0cm]{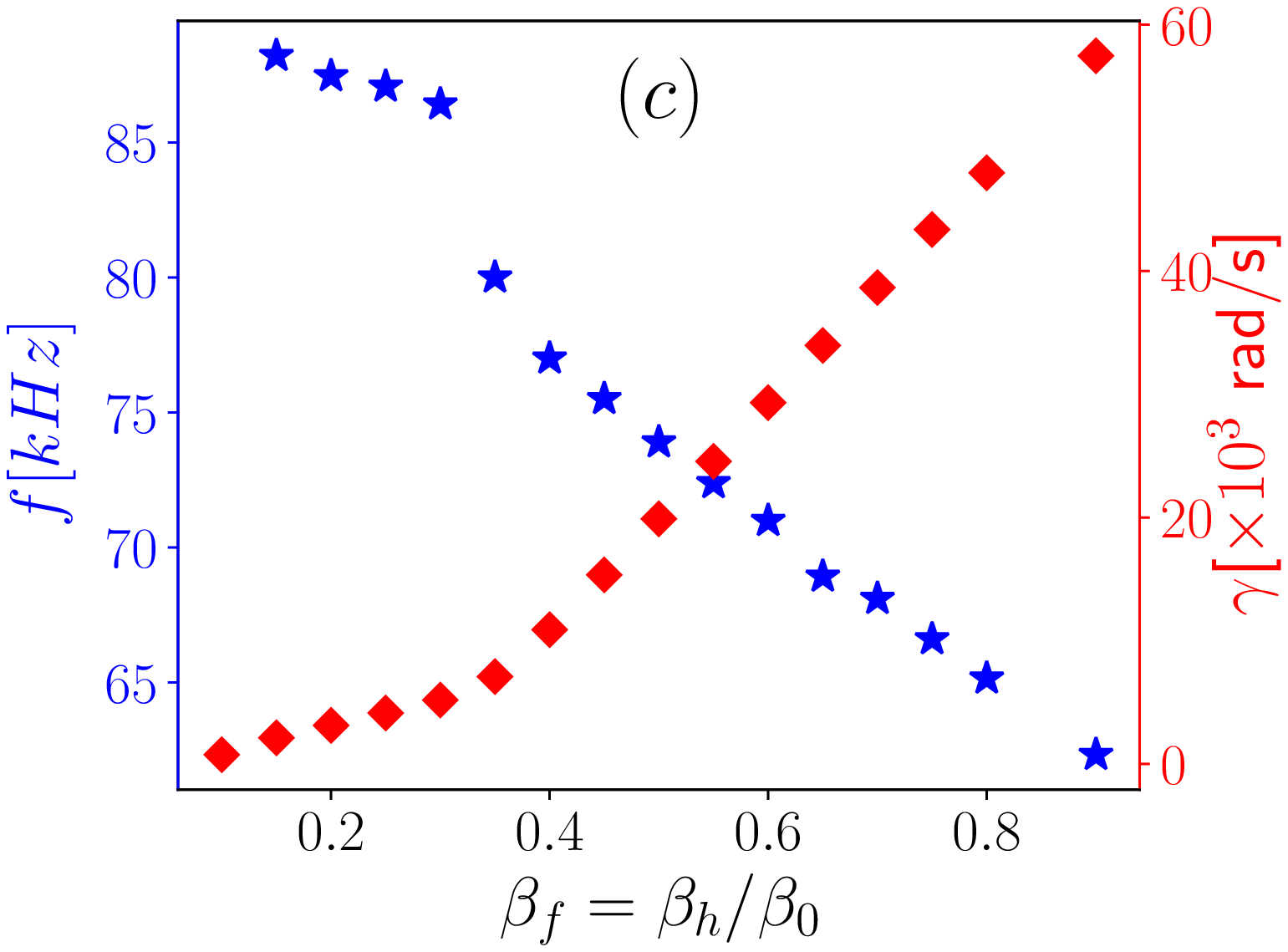}
  \caption{\label{fig:fgq0sn1} The frequency and growth rate dependence on
    $\beta_f=\beta_h/\beta_0$ with (a) $q_0=0.77$, (b) $q_0=0.9$, and
    (c) $q_0=1.05$.}
\end{figure}
For cases with $q_0=0.77$, as the EP fraction $\beta_f=\beta_h/\beta_0$
increases, the growth rate decreases first before rising again,
whereas the real frequency grows about linearly in both regimes.
The transition between the two regimes occurs at  $\beta_f=0.57$.
When the EP pressure is in the relatively low regime, the stabilizing effect of
EP on the kink mode is dominant; in the relatively high EP pressure regime,
fishbone mode can be excited. These results are consistent with previous
theories\cite{white1989high, wu1994alpha} and
simulations\cite{Kim2008Impact, fu2006global}.
When $q_0=0.90$,  the growth rate has the similar dependence on
$\beta_h/\beta_0$, with a lower transitional threshold $\beta_h/\beta_0=0.4$.
However, in higher $\beta_f$ regime, a new mode branch appears
where the real frequency is distinctively higher and decreases with
$\beta_f$. For $q_0=1.05$, only the new mode branch persists, where the growth
rate increases and the real frequency decreases with $\beta_f$. \par
\begin{figure}[ht]
  \centering
  \includegraphics[width=5.0cm]{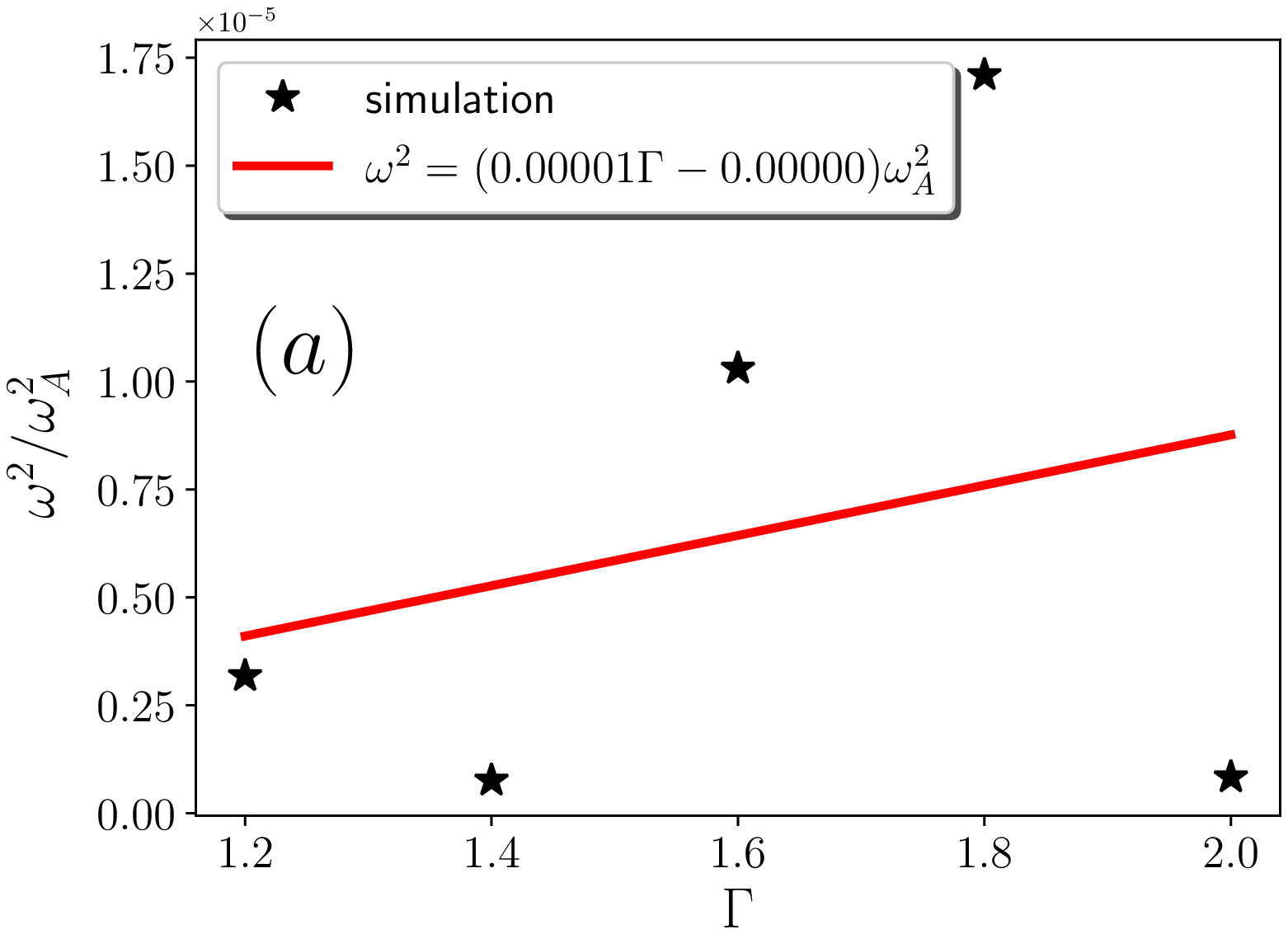}
  \includegraphics[width=5.0cm]{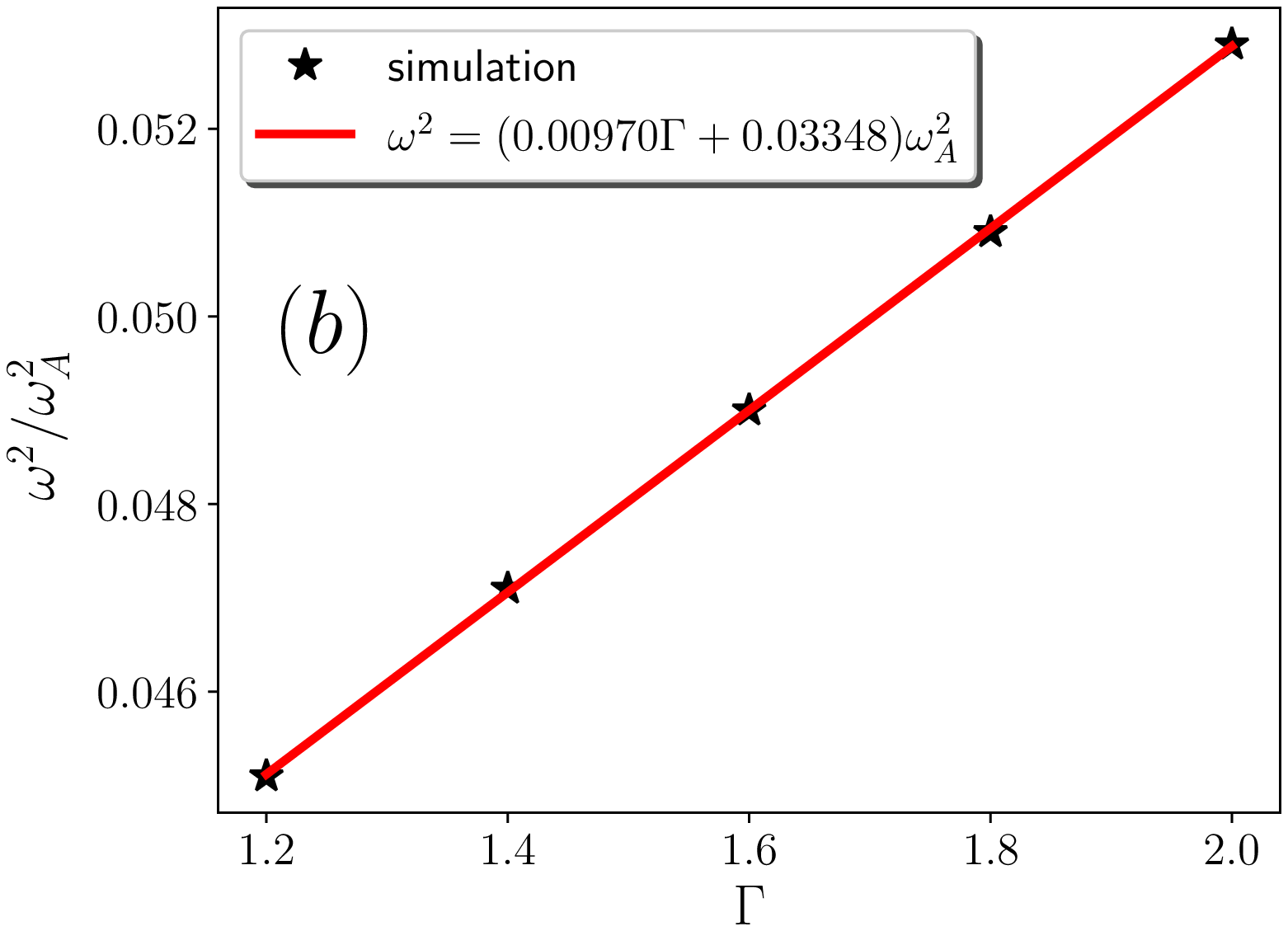}
  \includegraphics[width=5.0cm]{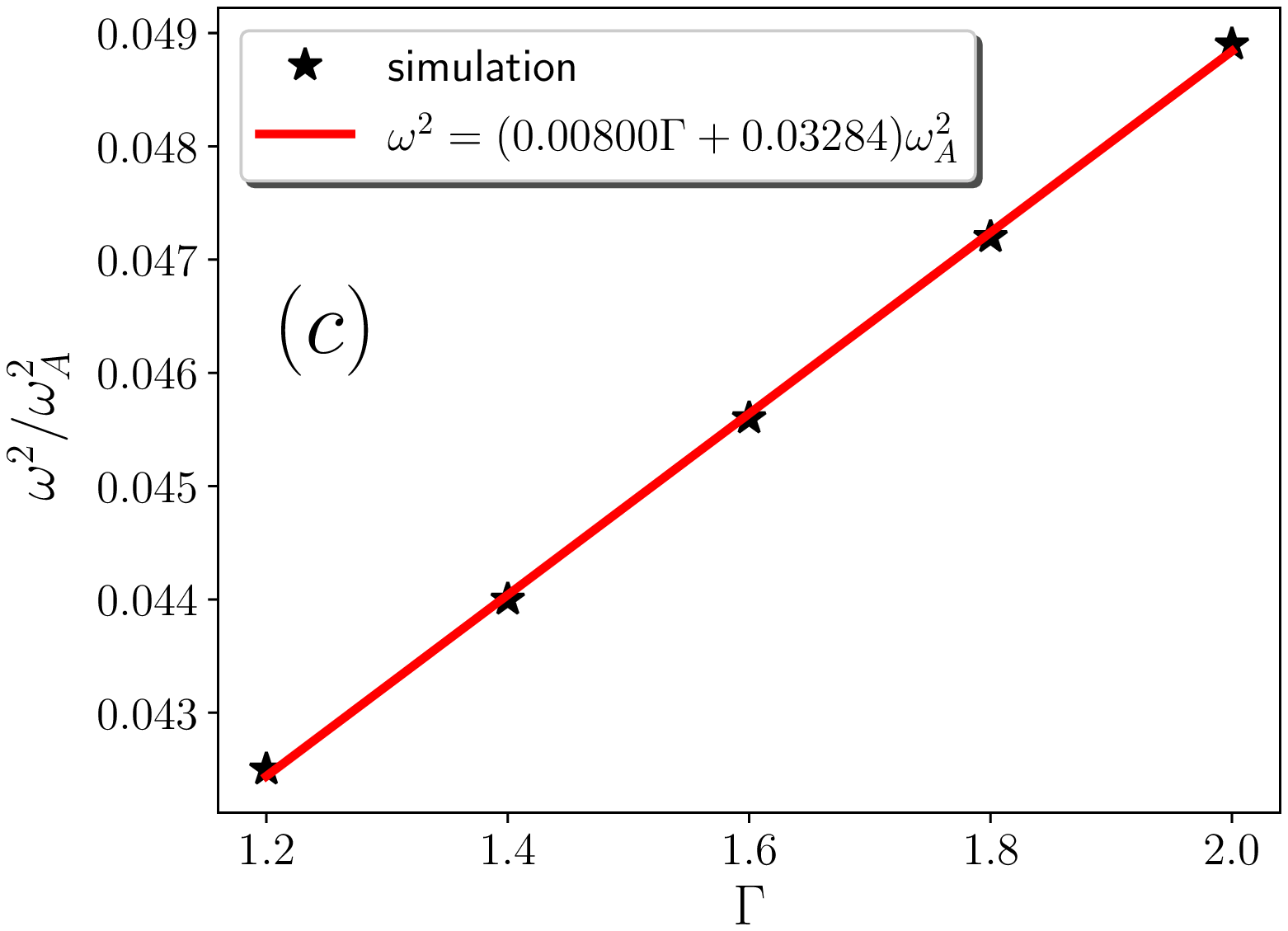}
  \caption{\label{fig:w2vssp} The square of normalized mode frequency
    as a function of the specific heat coefficient $\Gamma$ for
    (a) $q_0=0.77$; (b) $q_0=0.9$; and (c) $q_0=1.05$.}
\end{figure}
\begin{figure}[ht]
  \centering
  \includegraphics[width=5.0cm]{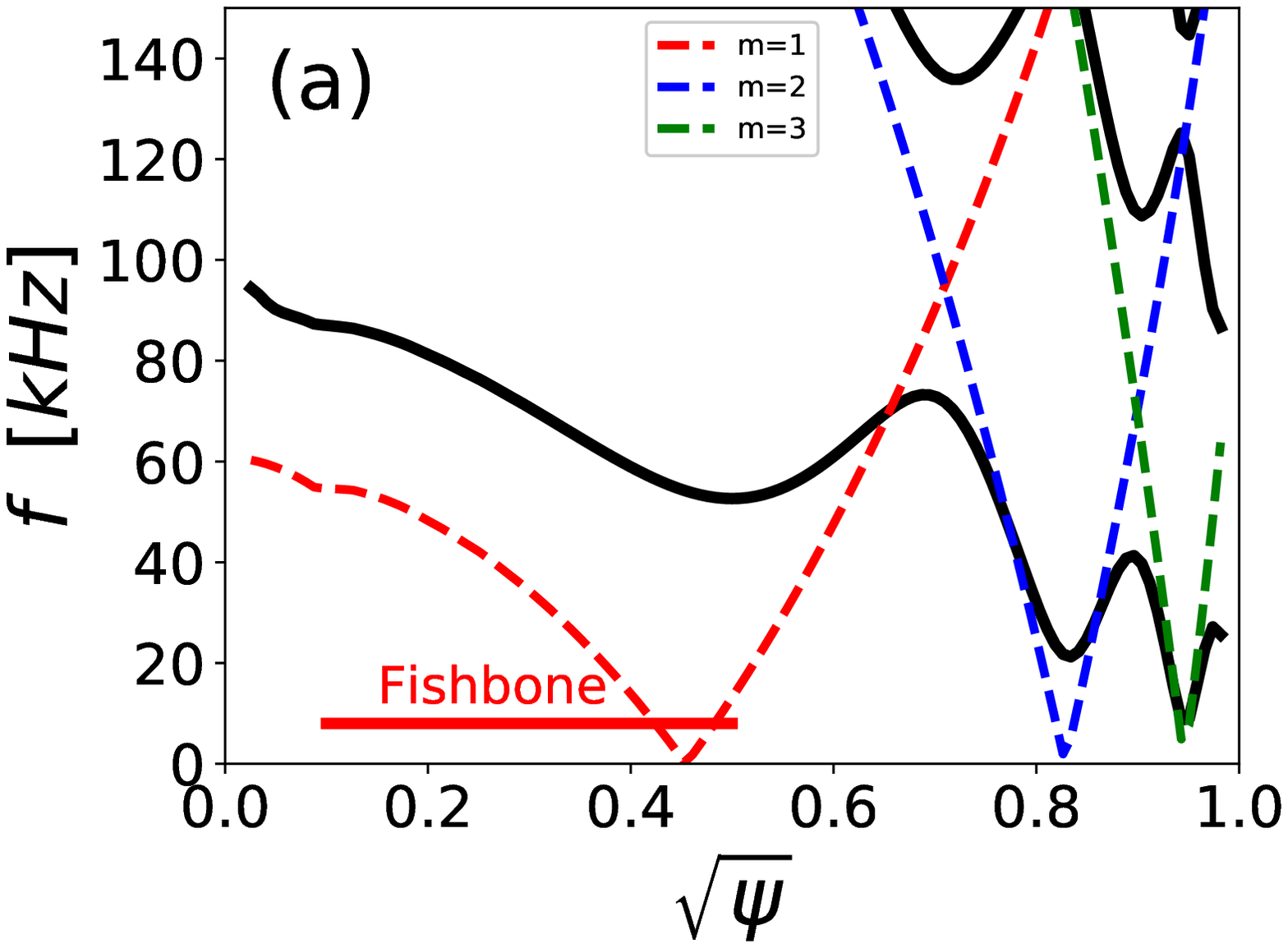}
  \includegraphics[width=5.0cm]{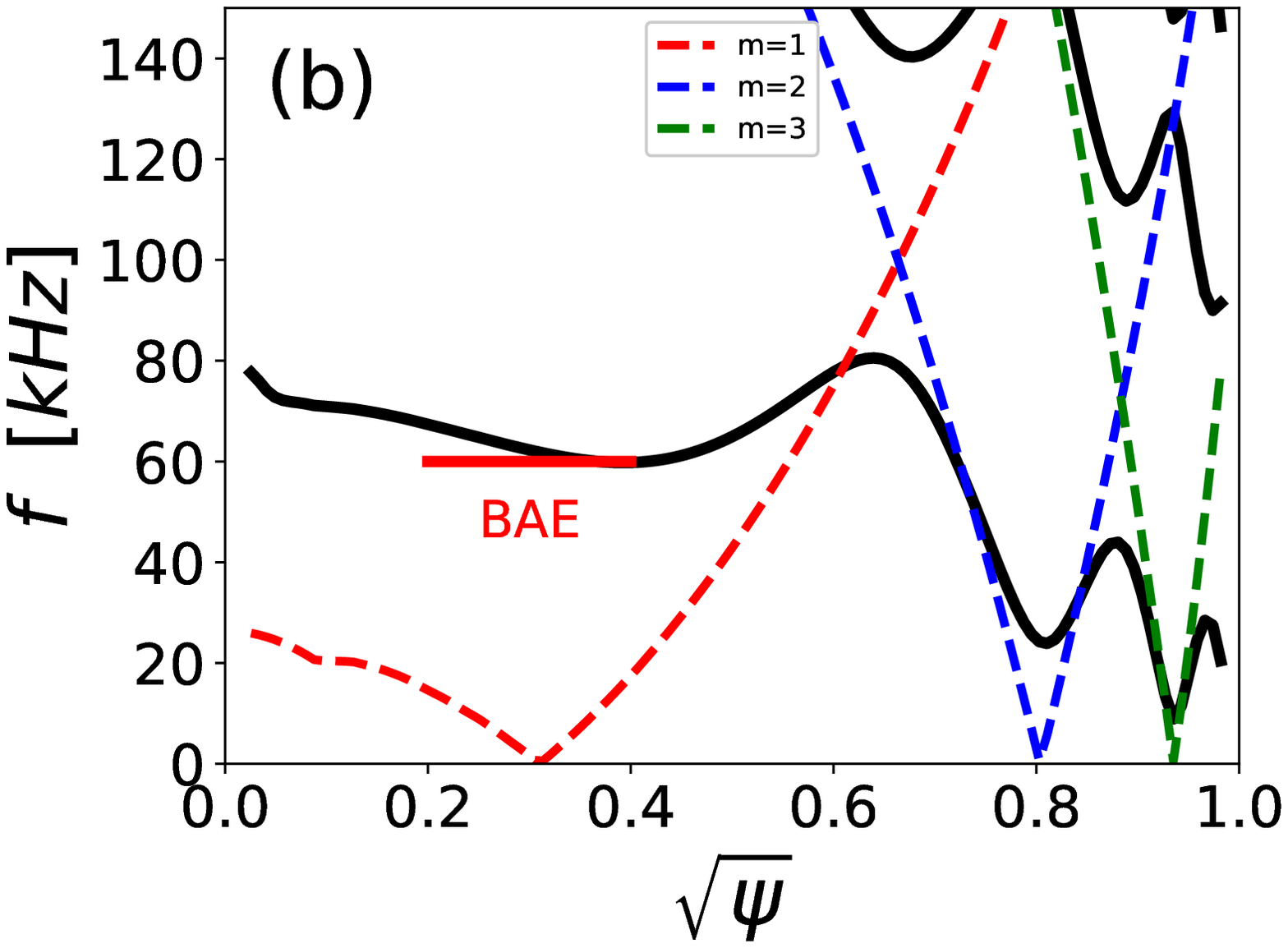}
  \includegraphics[width=5.0cm]{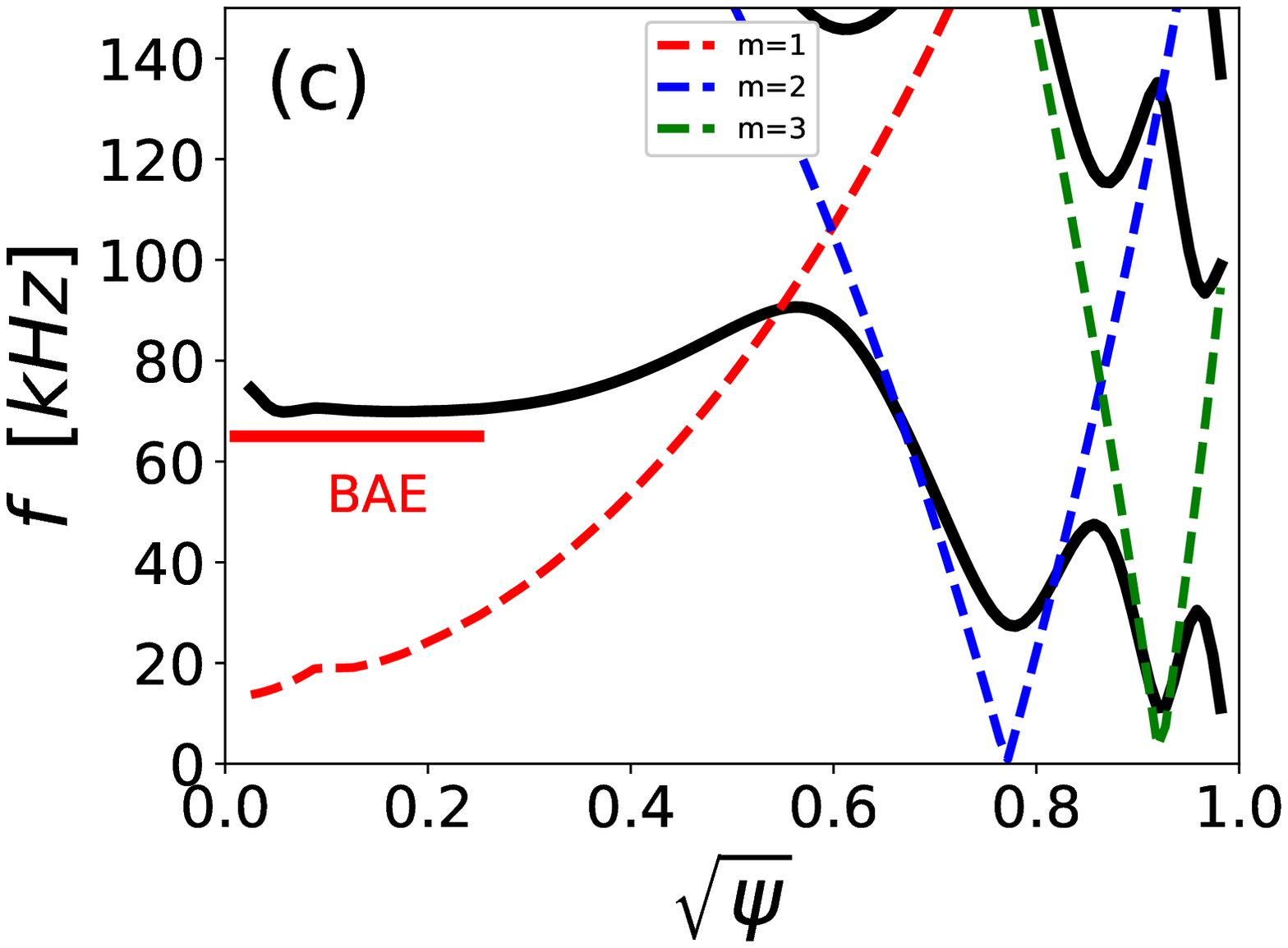}
  \caption{\label{fig:mfs} Alfv\'en continua with toroidal mode
    number $n = 1$ calculated based on the slow-sound approximation from
    AWEAC. Alfv\'en continua from the cylindrical geometry limit
    (dotted line) are also given. (a) $q_0=0.77$; (b) $q_0=0.9$; and
    (c) $q_0=1.05$.}
\end{figure}

In order to identify the nature of the higher frequency mode,
we vary the specific heat coefficient $\Gamma$ of the bulk plasma,
and calculate the growth rate. From FIG. \ref{fig:w2vssp} (b), we can see that,
when $q_0=0.9$ and $\beta_h/\beta_0=0.8$, the square of mode frequency
increases linearly with $\Gamma$, which is consistent with the property of
BAE\cite{shen2017hybrid}. Similarly, the BAE nature of the is also
verified for  $q_0=1.05$ and $\beta_h/\beta_0=0.8$ (FIG. \ref{fig:w2vssp}c).
In contrast, for $q_0=0.77$ and $\beta_h/\beta_0=0.8$,
no apparent relation between the mode frequency and $\Gamma$
can be found, where the fishbone mode dominates (FIG. \ref{fig:w2vssp}a).
Based on the Alfv\'en continua for toroidal mode number $n = 1$ calculated
using the  AWEAC code, the radial locations and frequencies of
the modes with $q_0=0.9$ and $q_0=1.05$ are well within the BAE
gap (FIG. \ref{fig:mfs}). \par
\begin{figure}[ht]
  \centering
  \begin{overpic}[scale=0.4]{figures/n1q76bh25p.eps}
    \put(13, 65){$(a)$}
  \end{overpic}
  \begin{overpic}[scale=0.2]{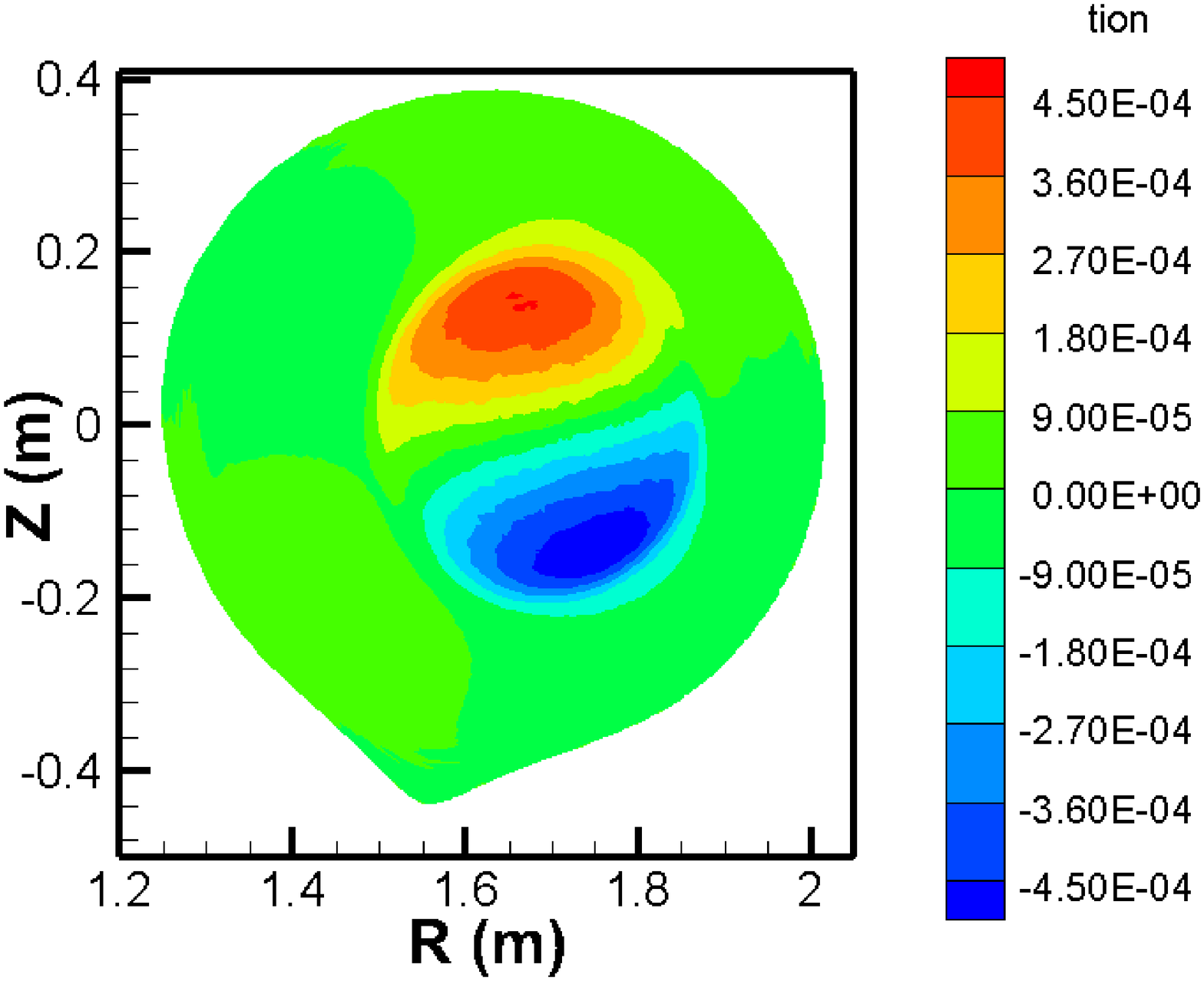}
    \put(13, 65){$(b)$}
  \end{overpic}
  \begin{overpic}[scale=0.2]{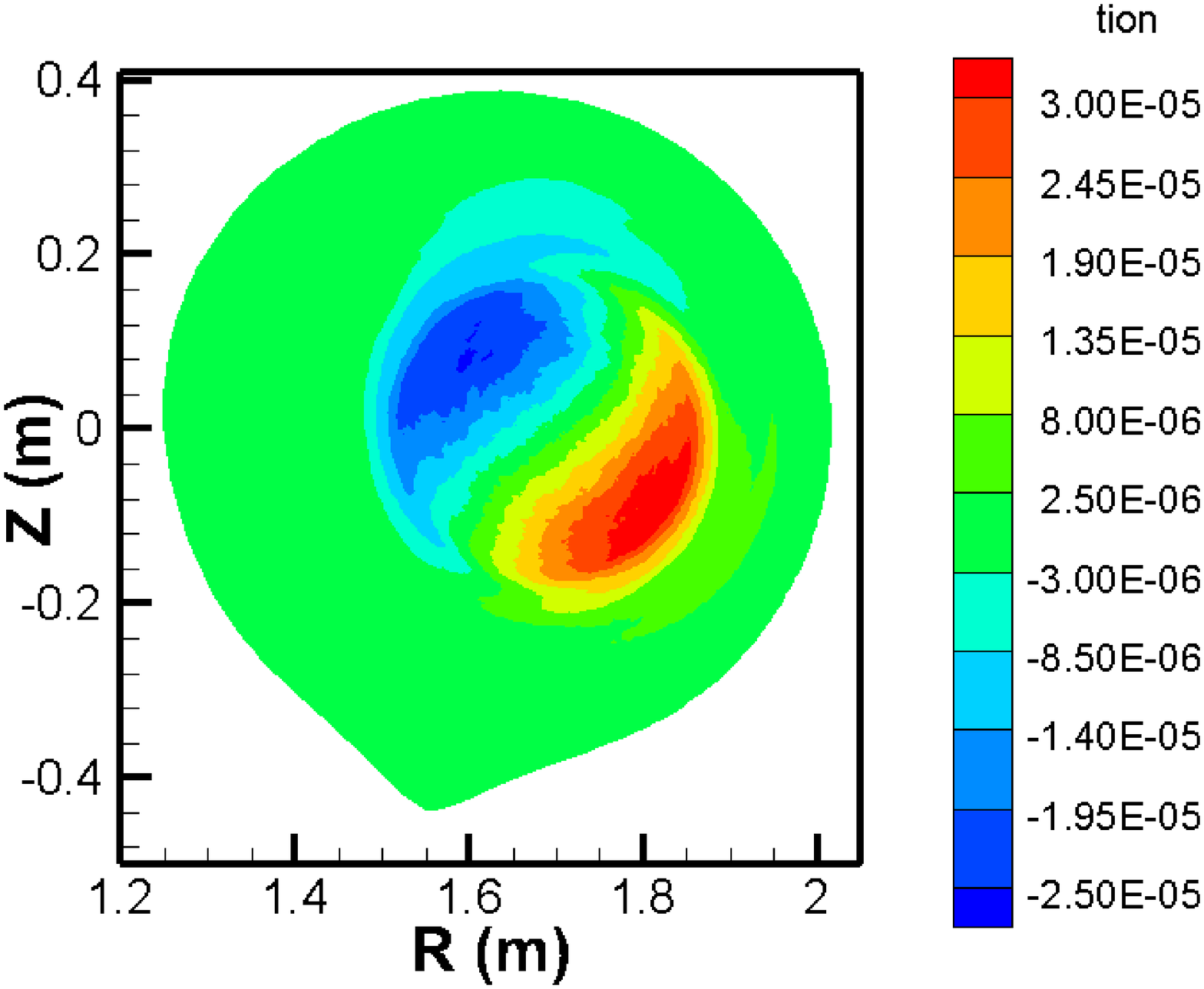}
    \put(13, 65){$(c)$}
  \end{overpic}
  \begin{overpic}[scale=0.24]{figures/q90beta0n1p.eps}
    \put(13, 65){$(d)$}
  \end{overpic}
  \begin{overpic}[scale=0.24]{figures/q90beta25n1p.eps}
    \put(13, 65){$(e)$}
  \end{overpic}
  \begin{overpic}[scale=0.24]{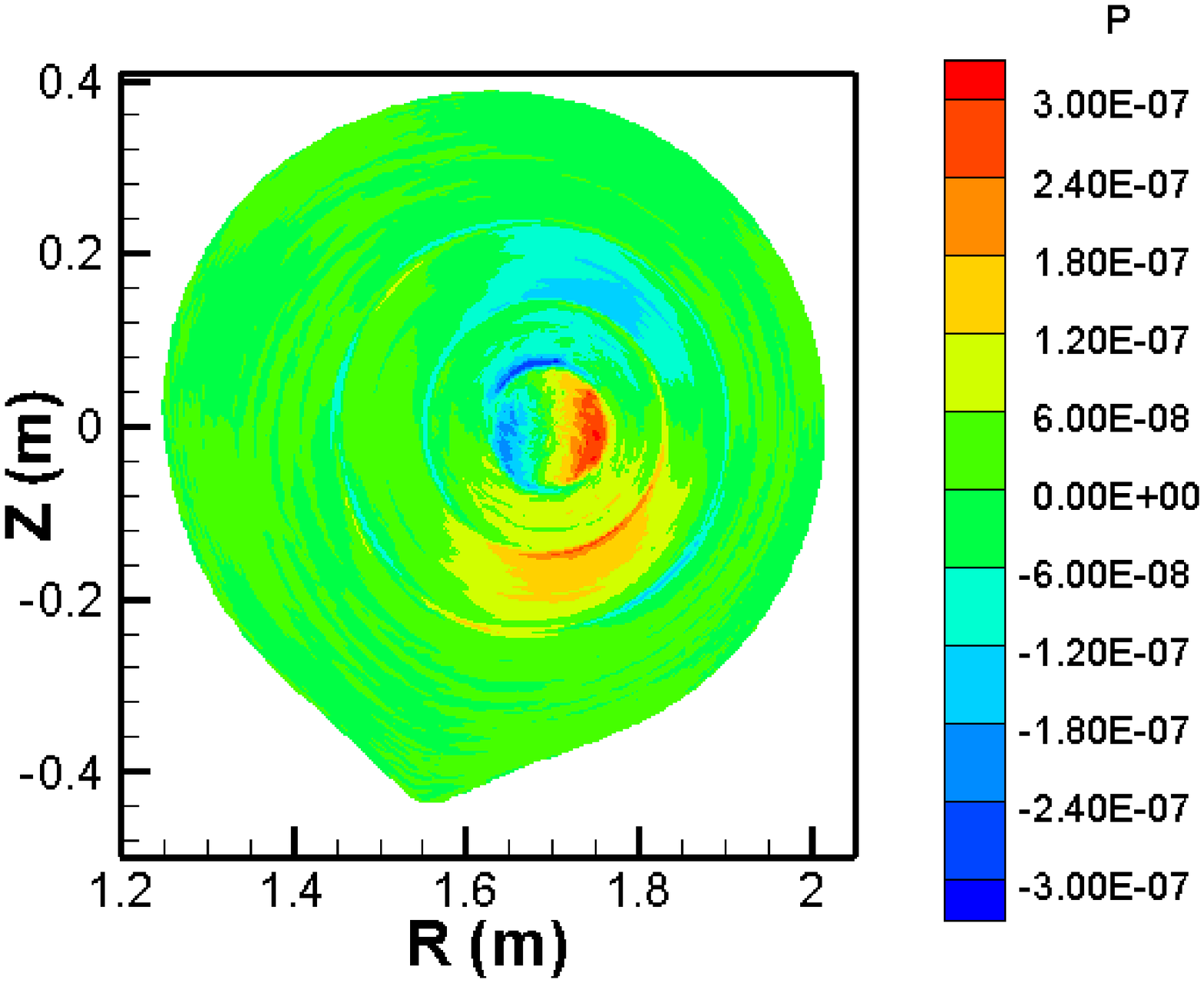}
    \put(13, 65){$(f)$}
  \end{overpic}
  \begin{overpic}[scale=0.24]{figures/q97beta25n1p.eps}
    \put(13, 65){$(g)$}
  \end{overpic}
  \begin{overpic}[scale=0.24]{figures/q102beta25n1p.eps}
    \put(13, 65){$(h)$}
  \end{overpic}
  \begin{overpic}[scale=0.33]{figures/q105beta25n1p.eps}
    \put(13, 65){$(i)$}
  \end{overpic}
  \caption{\label{fig:qbetahsn1} Contour plot of plasma pressure with
    (a) $q_0=0.77$, $\beta_h/\beta_0=0.25$,
    (b) $q_0=0.77$, $\beta_h/\beta_0=0.5$,
    (c) $q_0=0.77$, $\beta_h/\beta_0=0.75$,
    (d) $q_0=0.9$, $\beta_h/\beta_0=0.25$,
    (e) $q_0=0.9$, $\beta_h/\beta_0=0.50$,
    (f) $q_0=0.9$, $\beta_h/\beta_0=0.75$,
    (g) $q_0=1.05$, $\beta_h/\beta_0=0.25$,
    (h) $q_0=1.05$, $\beta_h/\beta_0=0.5$, and
    (i) $q_0=1.05$, $\beta_h/\beta_0=0.75$.}
\end{figure}

As both $q_0$ and $\beta_f$ increase from the kink dominant regime,
the transition to BAE dominant regime is also apparent in the variation
of mode structure (FIG. \ref{fig:qbetahsn1}). In the lower $q_0$ and
$\beta_f$ regime, the well-defined $(1, 1)$ kink mode structure
is localized inside the $q=1$ surface (FIG. \ref{fig:qbetahsn1}a).
This contrasts with the multiple-surface coupled mode structure for
BAE in the higher $q_0$ and $\beta_f$ regime (FIG. \ref{fig:qbetahsn1}i).
In between, the transition manifests in the mixture of the
characteristics of both kink and AE modes (e.g. FIG. \ref{fig:qbetahsn1}f).\par
\section{Conclusions}
In summary, the transition of the dominant $n=1$ mode from fishbone
to BAE instability in the HL-2A tokamak configuration has been
observed to take place when both the safety factor $q_0$ at
magnetic axis and the energetic particle fraction are above certain
threshold in hybrid-kinetic MHD simulations using
the NIMROD code. When $q_0$ is well below unity, the dominant
EP-driven instability is the fishbone mode; when $q_0$ is slightly
above unity, the dominant EP-driven instability becomes TAE and BAE
as the EP fraction increases. The transition occurs in between
these two regimes, where the mode frequency experiences abrupt jump
even though the mode growth rate varies continuously, and the mode
structure shows a mixture of signatures from both kink-fishbone and
TAE/BAE. These findings may help the identification and control of
the dominant EP-driven modes in experiments.\par
For the HL-2A experimental equilibrium we study in this work,
the $q$ profile is rather flat in the core region. The
effects of the weak magnetic shear on the EP-driven modes
remain to be better understood. In addition, as the EP fraction
of plasma $\beta$ increases, the finite orbit size of EP may no
longer be ignored. We plan on further addressing these issues in future work.

\section*{Acknowledgments}
This work was supported by the National Magnetic Confinement Fusion
Program of China Grant No. 2019YFE03050004,
the National Natural Science Foundation of China Grant Nos. 11875253,
11775221, 51821005, the Fundamental Research Funds for the Central
Universities Grant Nos. WK3420000004 and 2019kfyXJJS193,
the Collaborative Innovation Program of Hefei Science Center, CAS
Grant No. 2019HSC-CIP015, the U.S. Department of Energy
Grant Nos. DE-FG02-86ER53218 and DE-SC0018001.
This research used the computing resources from the Supercomputing
Center of University of Science and Technology of China.


\begin{thebibliography}{99}

\bibitem{Shafranov1970Hydromagnetic}%
  Shafranov V D
  1970
  \textit{Soviet Physics Technical Physics}
  \textbf{15} 175
\bibitem{Rosenbluth1973Nonlinear}%
  Rosenbluth M N
  \textit{et al} 1973
  \textit{Physics of Fluids}
  \textbf{16} 1894
\bibitem{Bussac1975Internal}%
  Bussac M N
  \textit{et al} 1975
  \textit{Physical Review Letters}
  \textbf{35} 1638
\bibitem{Fasoli2007}%
  Fasoli A
  \textit{et al} 2007
  \textit{Nuclear Fusion}
  \textbf{47} S264
\bibitem{Heidbrink1990}%
  Heidbrink W W and Sager G
  1990
  \textit{Nuclear Fusion}
  \textbf{30} 1015
\bibitem{Nave1991}%
  Nave M
  \textit{et al} 1991
  \textit{Nuclear Fusion}
  \textbf{31} 697
\bibitem{Chen2010}%
  Chen W
  \textit{et al} 2010
  \textit{Nuclear Fusion}
  \textbf{50} 084008
\bibitem{Xu2015}%
  Xu L Q
  \textit{et al} 2015
  \textit{Physics of Plasmas}
  \textbf{22} 122510
\bibitem{von1974studies}%
  Von Goeler S
  \textit{et al} 1974
  \textit{Physical Review Letters}
  \textbf{33} 1201
\bibitem{mcguire1983study}%
  McGuire K
  \textit{et al} 1983
  \textit{Physical Review Letters}
  \textbf{50} 891
\bibitem{chen1984excitation}%
  Chen L
  \textit{et al} 1984
  \textit{Physical Review Letters}
  \textbf{52} 1122
\bibitem{coppi1986theoretical}%
  Coppi B and Porcelli F
  1986
  \textit{Physical Review Letters}
  \textbf{57} 2272
\bibitem{Rosenbluth1975Excitation}%
  Rosenbluth M N and Rutherford P H
  1975
  \textit{Physical Review Letters}
  \textbf{34} 1428
\bibitem{Tsang1981Destabilitization}%
  Tsang K T
  \textit{et al} 1981
  \textit{The Physics of Fluids}
  \textbf{24} 1508
\bibitem{Fu1989Excitation}%
  Fu G Y and Van Dam J W
  1989
  \textit{Physics of Fluids B: Plasma Physics}
  \textbf{1} 1949
\bibitem{Fu1989Stability}%
  Fu G Y and Van Dam J W
  1989
  \textit{Physics of Fluids B: Plasma Physics}
  \textbf{1} 2404
\bibitem{Chen2016Physics}%
  Chen L and Zonca F
  2016
  \textit{Reviews of Modern Physics}
  \textbf{88} 015008
\bibitem{white1989high}%
  White R B
  \textit{et al} 1989
  \textit{Physical Review Letters}
  \textbf{62} 539
\bibitem{Cheng1986Low}%
  Cheng C Z and Chance M S
  1986
  \textit{The Physics of Fluids}
  \textbf{29} 3695
\bibitem{Heidbrink1993Observation}%
  Heidbrink W W
  \textit{et al} 1993
  \textit{Physical Review Letters}
  \textbf{71} 855
\bibitem{shen2017hybrid}%
  Shen W
  \textit{et al} 2017
  \textit{Nuclear Fusion}
  \textbf{57} 116035
\bibitem{Chen2018Stabilization}%
  Chen W
  \textit{et al} 2018
  \textit{Nuclear Fusion}
  \textbf{58} 014001
\bibitem{YuLM2013frequency}%
  Yu L M
  \textit{et al} 2013
  \textit{Nuclear Fusion}
  \textbf{53} 053002
\bibitem{YuLM2018toroidal}%
  Yu L M
  \textit{et al} 2018
  \textit{Physics of Plasmas}
  \textbf{25} 012112
\bibitem{Shi2019Beta}%
  Shi P W
  \textit{et al} 2019
  \textit{Nuclear Fusion}
  \textbf{59} 066015
\bibitem{DingXT2018}%
  Ding X T and Chen W
  2018
  \textit{Plasma Science and Technology}
  \textbf{20} 094008
\bibitem{sovinec2004nonlinear}%
  Sovinec C R
  \textit{et al} 2004
  \textit{Journal of Computational Physics}
  \textbf{195} 355
\bibitem{Kim2008Impact}%
  Kim C C and the NIMROD team
  2008
  \textit{Physics of Plasmas}
  \textbf{15} 072507
\bibitem{Hou2019AE}%
  Hou Y W
  \textit{et al} 2019
  \textit{Physics of Plasmas}
  \textbf{26} 082505
\bibitem{Goldston00}%
  Goldston R J and Rutherford P H
  \textit{Introduction to Plasma Physics} 2000
  (Institute of Physics, Philadelphia)
\bibitem{DengWei2014investigation}%
  Deng W
  \textit{et al} 2014
  \textit{Nuclear Fusion}
  \textbf{54} 013010
\bibitem{Zhang2014Long}%
  Zhang R B
  \textit{et al} 2014
  \textit{Plasma Physics $\&$ Controlled Fusion}
  \textbf{56} 095007
\bibitem{wu1994alpha}%
  Wu Y L
  \textit{et al} 1994
  \textit{Physics of Plasmas}
  \textbf{1} 3369
\bibitem{fu2006global}%
  Fu G Y
  \textit{et al} 2006
  \textit{Physics of Plasmas}
  \textbf{13} 052517
\end{thebibliography}
\end{document}